\documentclass[12pt]{article}
\usepackage{arxiv}

\usepackage{amssymb}
\usepackage[english]{babel}
\usepackage{lipsum}
\usepackage[colorlinks]{hyperref}
\usepackage{graphicx}
\usepackage{amsmath}
\usepackage{amssymb}
\usepackage[dvipsnames]{xcolor}
\usepackage[justification=centering]{caption}
\usepackage{subcaption}
\usepackage{here}
\usepackage{gensymb}
\usepackage{mathrsfs}
\usepackage{multirow}
\usepackage{textcomp}
\usepackage{stmaryrd}
\usepackage{siunitx}
\usepackage{colortbl}
\usepackage[sort&compress,square,numbers]{natbib}

 \title{Comparison of grain growth mean-field models regarding predicted grain size distributions}
 
 \renewcommand{\headeright}{Preprint}
\renewcommand{\undertitle}{Preprint}
\renewcommand{\shorttitle}{Comparison of grain growth mean-field models}

\date{September 15, 2023}

\author{ Marion Roth\thanks{Corresponding author.} \\
	Mines Paris\\
	PSL University\\
 Centre for Material Forming (CEMEF), UMR CNRS\\
	06904, Sophia Antipolis, France \\
	\texttt{marion.roth@minesparis.psl.eu} \\
	\And
	\href{https://orcid.org/0000-0001-6804-1974}{\includegraphics[scale=0.06]{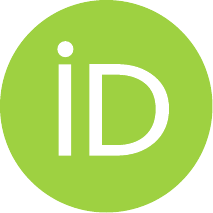}\hspace{1mm}Baptiste Flipon} \\
  	Mines Paris\\
	PSL University\\
 Centre for Material Forming (CEMEF), UMR CNRS\\
	06904, Sophia Antipolis, France \\
	\texttt{baptiste.flipon@minesparis.psl.eu} \\
 	\And
\href{https://orcid.org/0000-0002-8963-977X}{\includegraphics[scale=0.06]{orcid.pdf}\hspace{1mm}Nathalie Bozzolo} \\
	Mines Paris\\
	PSL University\\
 Centre for Material Forming (CEMEF), UMR CNRS\\
	06904, Sophia Antipolis, France \\
	\texttt{nathalie.bozzolo@minesparis.psl.eu} \\
 	\And
	\href{https://orcid.org/0000-0002-6677-2850}{\includegraphics[scale=0.06]{orcid.pdf}\hspace{1mm}Marc Bernacki} \\
	Mines Paris\\
	PSL University\\
 Centre for Material Forming (CEMEF), UMR CNRS\\
	06904, Sophia Antipolis, France \\
	\texttt{marc.bernacki@minesparis.psl.eu} \\
}

\begin{document}
\maketitle

\begin{abstract}

Mean-field models have the ability to predict grain size distribution evolution occurring through thermomechanical solicitations. This article focuses on a comparison of mean-field models under grain growth conditions. Different microstructure representations are considered and discussed, especially regarding the consideration of topology in the neighborhood construction. Experimental data obtained with a heat treatment campaign on a 316L austenitic stainless steel are used for material parameters identification and as a reference for model comparisons. Mean-field models are also confronted to both mono- and bimodal initial grain size distributions to investigate the interest of introducing neighborhood topology in microstructure predictions models. This article exposes that improvements in the predictions are obtained in monomodal cases for topological models. In bimodal test, no comparison with experimental data were performed as no data were available. But relative comparisons between models indicate few differences in predictions. The interest of neighborhood topology in grain growth mean-field models gives overall small improvements compared to classical mean-field models when comparing implementation complexity.

\end{abstract}


\section{Introduction}
	
The phenomenon of grain growth takes place in metallic materials when their are submitted to a heat treatment. Considered as the only phenomenon occurring, materials are assumed free of any stored energy (\textit{i.e.} low dislocation density) and the driving pressure for grain boundary (GB) migration arises from the minimization of the GB surface energy leading to a curvature flow problem. At the polycrystalline scale, the GB motion is generally described by $v = M_{GB} \left|P\right|$ with $v$ the velocity norm of the boundary \cite{Rollett2017}, $M_{GB}$ its mobility and $P=-\gamma_{GB}\kappa$ with $\gamma_{GB}$ the GB energy and $\kappa$ the trace of the curvature tensor.\\
		
Since 70 years, models have largely been developed in order to predict microstructure changes under thermomechanical treatments and their impact on macroscopic properties. At the polycrystalline scale, three types of models can be found in the literature: phenomenological, mean-field, and full-field models. Phenomenological approaches are classically based on an experimental database and correspond to the extraction of a mathematical trend from what is observed experimentally. Such models are restricted to the set of thermomechanical conditions and to the material experimentally investigated \cite{Avrami1939,JohnsonMehl1939}. The last two model types are based on a more generic approach, since they use physical equations to predict microstructure evolution. Full-field models, at the mesoscopic scale, give access to a complete description of the system where each individual grain and its topology are taken into account \cite{Maire2017,bernacki:hal-00509731,Hallberg2011}. It has the advantage to be able to deal with local heterogeneities. However, their major drawback is to be very costly in terms of computing time. \\

Mean-field models, on the other hand, keeps this physically based implementation with a similar set of equations, but the general description of the microstructure is simplified \cite{Hillert1965,Abbruzzese1992,Montheillet2009,Cram2009,Bernard2011,Favre2013,Maire2018}. They also present competitive computation time when compared to full-field models. The original definition of a mean-field model considers the evolution of a microstructure described by mean quantities. The work of Burke and Turnbull \cite{BurkeTurnbull1952} provides the simplest version of such models. In this case, the microstructure is solely defined by its mean grain radius $\bar{R}$. The mean grain size (MGS) quantity is commonly determined by the (arithmetic) mean of the equivalent diameter ($\overline{ED}=2\bar{R}$) where $ED_i$ is defined in 2D, resp. in 3D, as the diameter of a circle, resp. a sphere, having the same area ($A_i$), resp. volume ($V_i$), as the considered grain $G_i$:
    \begin{equation}
         \bar{R}= \frac{\overline{ED}}{2}=\frac{1}{2N}\sum_{i=1}^{N}ED_i, \text{ i.e }\bar{R}= \frac{1}{N}\sum_{i=1}^{N}\left(\frac{A_i}{\pi}\right)^{1/2} \text { in 2D, and }  \bar{R}= \frac{1}{N}\sum_{i=1}^{N}\left(\frac{3V_i}{4\pi}\right)^{1/3} \text { in 3D,}
    \end{equation}
 with $N$ the number of grains in the microstructure. 

 In Burke and Turnbull (B\&T) context, the GB velocity norm is considered proportional to the inverse of $\bar{R}$. Hillert model \cite{Hillert1965} introduces the consideration of the grain size distribution (GSD) in mean-field models. The microstructure is sampled in a representative distribution, where each bin is assigned to a given ED value and a frequency. In the GSD, a bin is commonly called grain class, each of them being virtually composed of many grains indicated by the frequency. For every class, one can define a representative grain having the characteristics of the associated grain class.  Hillert formalism represents the microstructure as a grain embedded in a Homogenous Equivalent Medium (HEM) that is characterized by the MGS of the distribution. The size evolution of a considered class is deduced from the difference in curvature radius between the current class and that of the HEM. Later on, statistical models, as the one developed by Abbruzzese \textit{et al.} \cite{Abbruzzese1992,Lucke1992} modify the definition of the HEM. Rather than being only associated with a mean grain radius, the HEM is replaced by contact surfaces defined for all grain classes of the microstructure. Each grain class is surrounded by a statistical medium (SM) composed of all the grain classes of the microstructure. The contact surface between a grain class and its surrounding neighbors is defined by a perimeter intersection probability. One of the latest topological mean-field models, developed by Maire \textit{et al.} \cite{Maire2018} combines several formalisms in its neighborhood construction. This hybrid model uses both the statistical approach initiated by Abbruzzese \textit{et al.} and a deterministic number of neighbors ruled by the imposed bijectivity of neighborhood assignment. To the authors knowledge, the interest of these different views concerning the neighborhood description was never discussed in the state of the art for GG modeling. The main purpose of this work is to compare mean-field models of different microstructure descriptions to determine if semi-topological approaches are of interest to better describe GSD evolution in the context of grain growth. \\
 
This article is constructed as follows: first, mean-field GG models are introduced. A detailed description of the microstructure formalism in Hillert, Abbruzzese  \textit{et al.} and Maire  \textit{et al.} models is made. Then optimized data parameters for the use of these models are determined and discussed. The last section is dedicated to a distribution comparison of the Maire \textit{et al.} model to the mean-field models of Hillert, Abbruzzese \textit{et al.} and to experimental data.

	\section{Mean-field models} \label{MFM}
	
	 This section recalls the main equations and details the microstructure description of the previously introduced models for the GG phenomenon. \\

	\subsection{Burke and Turnbull model}
	
	B\&T model \cite{BurkeTurnbull1952} illustrates the original meaning of mean-field models as the microstructure description is reduced to mean geometric considerations. Indeed, the rate of GG ($d\bar{R}/dt$) is assumed proportional to $1/\bar{R}$ \cite{Rollett2017} such as: 
	\begin{equation}
		\label{BetT}
		d\bar{R}=M_{GB}P_c dt\text{, with } P_c = \frac{\alpha \left(d-1\right)\gamma_{GB}}{\bar{R}},
	\end{equation}
	where $\alpha$ is a constant, $\gamma_{GB}$ is the GB energy, $M_{GB}$ is the GB mobility, and $d$ the space dimension. In the B\&T analysis, $\gamma_{GB}$ and $M_{GB}$ are assumed to be constant in time and space.\\
	
	A parabolic expression can then be derived from eq. \eqref{BetT} and represents the time dependence of $\bar{R}^2 - {\bar{R}_0}^2$: 
		\begin{equation}
			\bar{R}^2 - {\bar{R}_0}^2 = 2\left(d-1\right)\alpha M_{GB} \gamma_{GB} t,
			\label{parab_law_BT}
		\end{equation}
	where $\bar{R}_0$ is the initial mean grain radius. An extension of this law is classically preferred in the literature : 
		\begin{equation}
		\bar{R}^2 - {\bar{R}_0}^2 = \tilde{\alpha} M_{GB} \gamma_{GB} t^n,
		\label{parab_law_BT2}
	\end{equation}
    with $n$ and $\tilde{\alpha}$ two constants to be fitted thanks to experimental data.
	
	\subsection{Hillert model}\label{Hillert}
	
\begin{figure}[h!]
	\centering
	\includegraphics[width=0.32\textwidth]{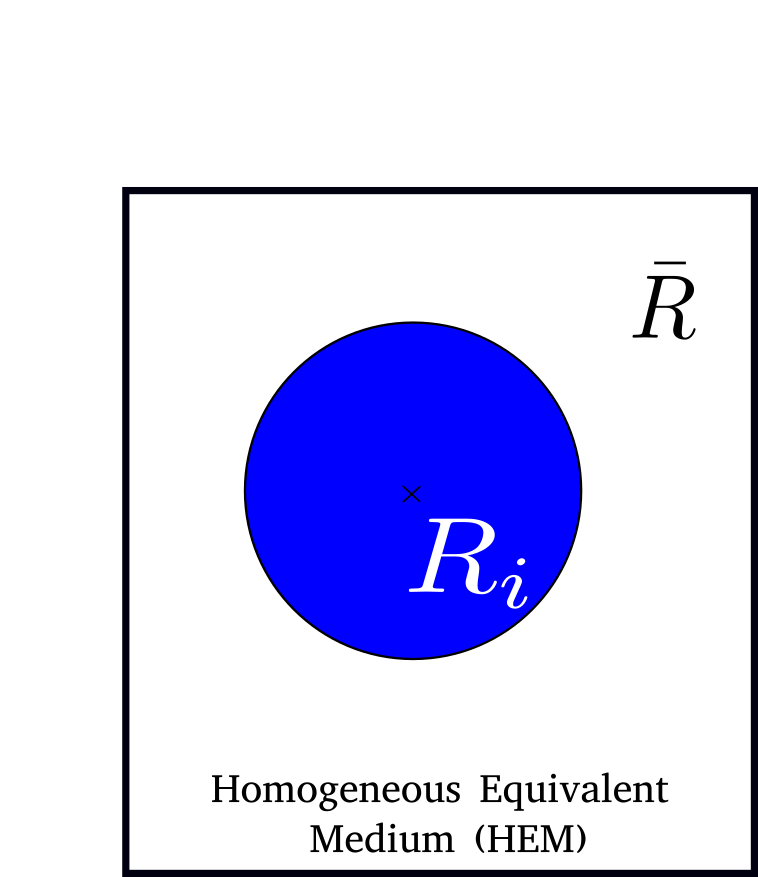}
	\caption{\label{fig:Hillert} Schematic representation of the microstructure in the Hillert model.}
    \end{figure}

	The GG mean-field model proposed by Hillert \cite{Hillert1965} relies on considering each grain class inside a common HEM. The microstructure description is illustrated in fig. \ref{fig:Hillert} where the representative grain of a given grain class $R_i$ is surrounded by a HEM defined by the mean grain radius of the microstructure $\bar{R}$. Compared to B\&T model, the driving pressure term is modified following a more local approach. $P_c$, for the class $i$, is defined as a function of the curvature radius of the grain class $i$: 
	 
	\begin{equation}
		dR_i = \frac{(d-1)}{2} M_{GB} \gamma_{GB}\left(\frac{1}{\bar{R}} -\frac{1}{R_i} \right)dt.
		\label{eq:Hillert}
	\end{equation}
	Thus, the grain $i$ will shrink when $\bar{R}>R_i$, will grow if $\bar{R}<R_i$ and will be stable when $\bar{R}=R_i$.
	
	\subsection{Abbruzzese \textit{et al.} model}\label{subsec:Abb_2D}
 
\begin{figure}[!h]
		\centering
		\begin{subfigure}[b]{0.32\textwidth}
			\includegraphics[width=\textwidth]{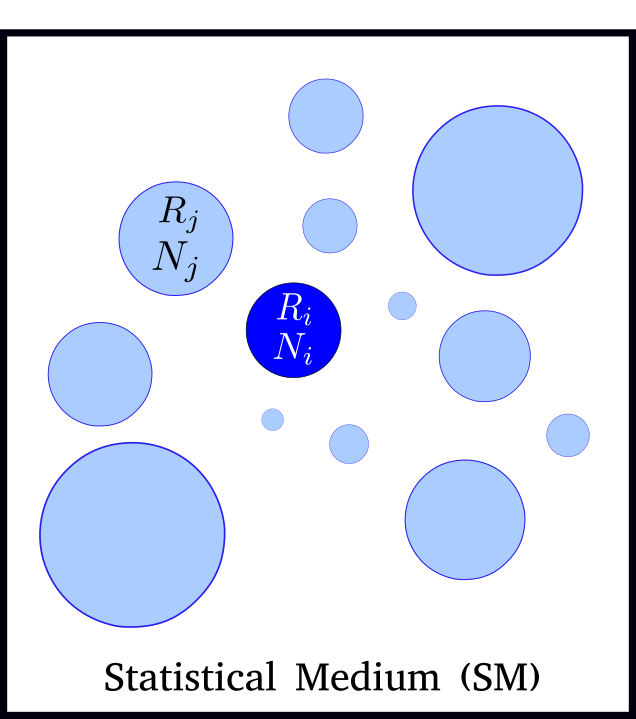}
			\caption{\label{fig:Abb_2Da}}
		\end{subfigure}
		\begin{subfigure}[b]{0.32\textwidth}
			\includegraphics[width=\textwidth]{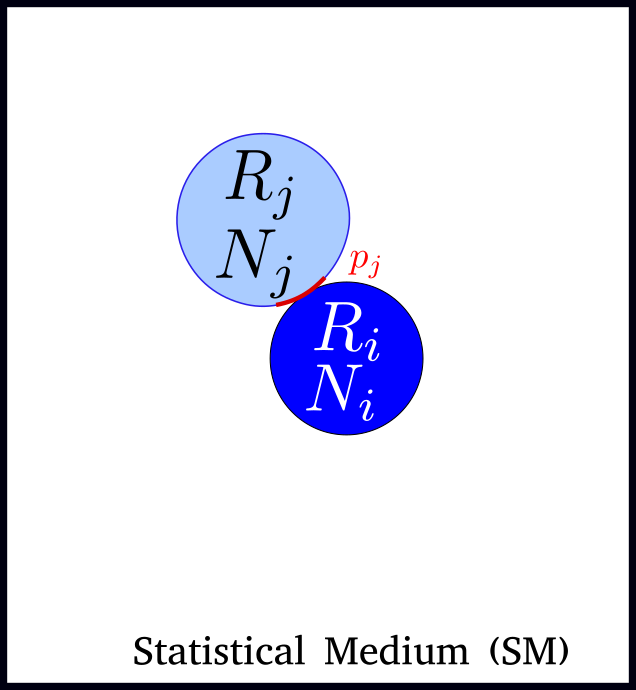}
			\caption{\label{fig:Abb_2Db}} 
		\end{subfigure}
		\caption{\label{fig:Abb_2D}Statistical neighborhood construction of the 2D GG model of Abbruzzese  \textit{et al.}: (a) description of the statistical medium, and (b) illustration of the contact probability $p_j$ concept.}
	\end{figure}

	In the statistical model proposed by Abbruzzese \textit{et al.} \cite{Abbruzzese1992, Lucke1992} in 1992, the authors suggested an evolution of the Hillert model by introducing a statistically constructed medium (SM) composed of all grain classes of the microstructure as illustrated in fig. \ref{fig:Abb_2Da}. 
	The contribution of all neighbors being weighted by a statistical coefficient: the contact probability.
	
	In 2D, considering the grain class $i$, the contact between $i$ and its neighbor grain classes is defined by a fraction of the neighbor class perimeter. The probability of a grain class $j$ to belong to the SM is only depending on its own size (grain radius -- $R_j$) and will be the same for all grain class $i$ considered. This represents what would be the chances or probabilities that a grain of the $j\textsuperscript{th}$ class in the microstructure intersects the representative grain $i$ as schematized in fig. \ref{fig:Abb_2Db}. The contact probability $p_j$ is then given by a ratio between the $j\textsuperscript{th}$ grain perimeter and the sum of all the grain perimeters of the microstructure. Along with the grain size, the number frequency of each grain class is also taken into account. $p_j$ expression is written as follows: 
	
	\begin{equation}
		p_j = \frac{N_j R_j}{\sum_{k=1}^{n}{N_k R_k}},
		\label{contact_prob_2D_abb}
	\end{equation}
	with $n$ the total number of grain classes in the microstructure and, $\forall k\in\llbracket 1,n\rrbracket$, $N_k$ the number of grains belonging to the grain class $k$.

	With such a kind of explicit description of the surrounding, the GB migration can be achieved locally between the class $i$ and each of its neighbors. In the context of 2D-GG ($d=2$), the difference of driving pressure between a grain class $i$ and its neighbor $j$ can be derived from the Hillert equation (eq. \eqref{eq:Hillert}) and is then given by:
	
	\begin{equation}
		dR_{(i,j)} = \frac{1}{2}M_{GB}\gamma_{GB} \left( \frac{1}{R_j} - \frac{1}{R_i} \right)dt.
		\label{eq:dRij_Abb}
	\end{equation} 
	
	The global variation of grain size undergone by the grain class $i$ is the sum of the local variations with every $j\textsuperscript{th}$ neighbor.
	The local GB migration can be generalized to all the $j\textsuperscript{th}$ neighbors of the grain class $i$ using the contact probability $p_j$. The total rate of evolution with respect to time for the radius of the representative grain $i$ can then be written as: 
	
	\begin{equation}
		dR_i = \sum_{j=1}^{n} p_j dR_{(i,j)}=\frac{1}{2}M_{GB}\gamma_{GB} \left(\sum_{j=1}^{n} \frac{p_j}{R_j} - \frac{1}{R_i} \right)dt.
		\label{eq Pij_2D}
	\end{equation}

	\subsection{Maire \textit{et al.} model} \label{Maire_model}
	
	Maire \textit{et al.} \cite{Maire2018} proposed a 3D mean-field model to simulate microstructure evolution under thermomechanical solicitations. Physical mechanisms such as GG, discontinuous dynamic recrystallization and post-dynamic recrystallization can be modeled. It is based on previous works of Bernard \textit{et al.} \cite{Bernard2011} and Beltran \textit{et al.} \cite{Beltran2015} who developed a mean-field model with two HEMs, one medium associated to non recrystallized grains and the other to recrystallized ones, each of these media being characterized by their MGS. Maire \textit{et al.} work focused on the introduction of neighborhood topology in the Bernard-Beltran previously defined formalism.
	A specific neighborhood is proposed for each grain class of the microstructure. Each class $i$ is characterized, in the context of GG, by two main properties: a grain radius $R_i$ and a number of grains belonging to this class $N_i$.
	Microstructure and its evolution are described as detailed below. The following equation links the radius variation $dR_{(i, j)}$ between two representative grains to the volume variation $dV_{(i,j)}$ associated to the GB migration: 
	\begin{equation}
		dV_{(i,j)} =  dR_{(i,j)} \times S_{c(i,j)} = dR_{(i,j)}  \times p_{(i,j)}\times S_{R_i},
		\label{eq variation volume}
	\end{equation} 
    \begin{equation}
        \text{with } dR_{(i,j)} = M_{GB}\gamma_{GB} \left( \frac{1}{R_j} - \frac{1}{R_i} \right)dt\text{, as } d=3 \text{ (see eq. \eqref{eq:Hillert})},
		\label{eq:Pij}
	\end{equation} 
	and, with $S_{c(i,j)}$ the contact surface between class $i$ and $j$ which is the product of $p_{(i,j)}$ the contact probability and $S_{R_i}$, the remaining surface of the grain class i:
	\begin{equation}
	     S_{c(i,j)} =  p_{(i,j)}\times S_{R_i}.
	\end{equation}


	\begin{figure}[!h]
		\centering
		\includegraphics[width=0.32\textwidth]{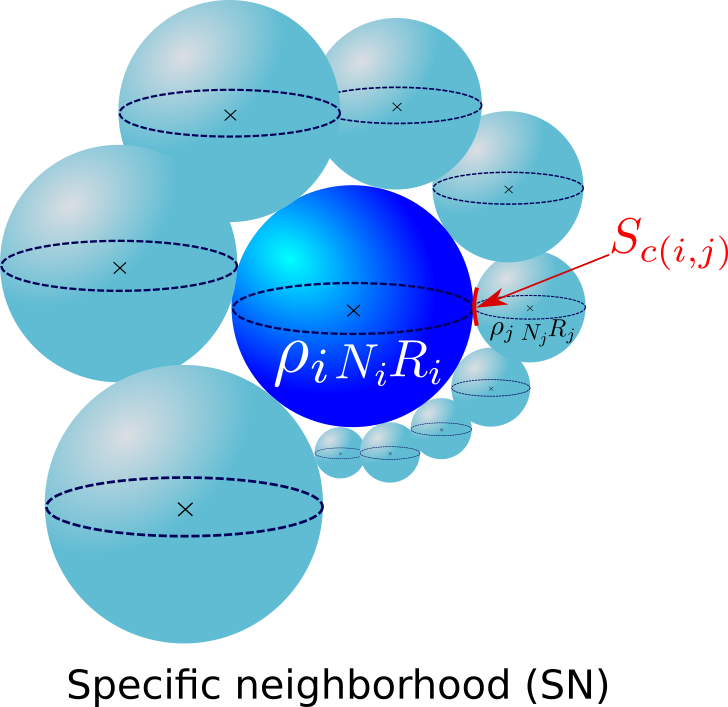}
		\caption{\label{fig:NHM}Microstructure description in the Maire \textit{et al.} model.}
	\end{figure}
	
	 Fig. \ref{fig:NHM} illustrates the microstructure description considered in the Maire \textit{et al.} model.

	This model considers a specific neighborhood (SN) that is built for each grain class based on two main criteria. First, bijectivity is ensured between two neighbor classes. For two considered neighbor grain classes $i$ and $j$, the contact probability of grain class $i$ with the grain class $j$ is computed as $p_{(i,j)}$ and it is automatically associated with the grain class $j$ as the contact probability $p_{(j,i)}=p_{(i,j)}$. The second criterion imposed by this model, intimately linked to the first one, is the volume conservation that will be defined in more details below. 
	
	%
	
	At each time increment of a grain growth simulation, the surface of the spherical representative grain of a class is filled with a neighborhood. Initially, each grain class has a remaining (available) surface equal to the surface of a sphere of radius $R_i(t=0)$: $S_{R_{i}} = 4 \pi R_{i}^2$. During the SN construction, the remaining available surface will decrease as neighbors are assigned until the SN of the considered grain class is completely filled (\textit{i.e.} no more available surface). 
	
	The filling of the SN is arbitrarily selected to be done in the ascending order of the GSD. By considering $i$ as the current class, a first neighbor class $i+1$ is attributed to the class $i$. The corresponding contact surface $S_{c{(i,i+1)}}$ is applied between the two considered grain classes and $S_{R_{i}}$ is updated as follows:
	\begin{equation}
		S_{R_{i}} =  4\pi R_i^2 -  S_{c{(i,i+1)}},
	\end{equation}
	$S_{c{(i,i+1)}}$ computation will be described with more detailed below.
	
	\begin{figure}[!h]
		\centering
		\includegraphics[width=0.4\textwidth]{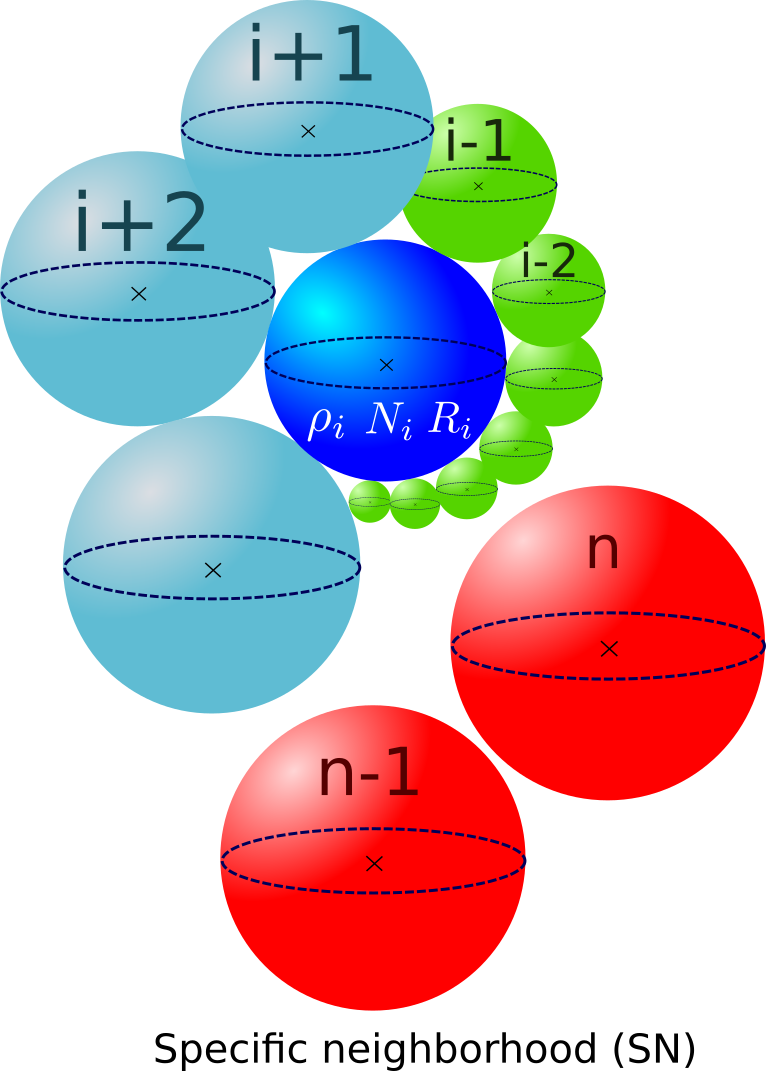}
		\caption{\label{Neighbor_construction}Specific Neighborhood construction in the Maire \textit{et al.} model for class $i$.}
	\end{figure}

	The next neighbors,  $j=i+2,\ i+3, \ldots$ for class $i$, are then selected following the ascending order of the initial GSD and a contact surface is one more time computed. This goes with all the microstructure grain classes with incomplete neighborhood. 
	Once the $i\textsuperscript{th}$ grain class neighborhood is filled. The following class, denoted $i+1$ here, becomes the current class and so on until the whole grain distribution is scanned. As the current class evolving in the GSD, a part of its neighborhood is already filled up by bijectivity with classes of smaller radius. In fig. \ref{Neighbor_construction}, the current class $i$ considered is taken in the middle of the GSD.  Green classes represent the ones attributed to the grain class $i$ by bijectivity when $i-1,i-2,\ldots$ were current grain classes. The blue ones are all the remaining classes of the distribution with no complete neighborhood. These classes are added to grain $i$ neighborhood when it is the current class. Finally, red grain classes do not belong to the grain class i neighborhood as their neighborhood are already complete entirely by bijectivity with smaller radius grain classes.

    As mentioned in the introduction, a bijective criterion is applied here. The remaining surface of the grain class $i+1$, $S_{R_{i+1}}$, has already been updated when $i+1$ has been attributed as the neighbor to the grain class $k\text{, with } k\leq i$. 
	That way, as the current class goes up in the distribution a part of the neighborhood has already been filled by bijectivity with grains having a radius smaller than the current class. After update, the remaining surface of grain class $j$ progressively decreases. A test is realized to check if  $S_{R_j} > 0$: if true, grain class $j$ can take new neighbors; and if not, grain class $j$ neighborhood is complete. Class $j$ is therefore not available anymore as a neighbor and will not take part in the construction of the following grain classes (class $j$ will be seen as the red grain classes from fig. \ref{Neighbor_construction} for all other classes)
 
    Of course, it is important here to highlight that the order used to build the neighborhood of each class can have an impact on the global answer of the GG model as the neighborhood determination will be different for example if the distribution is treated by a decreasing GSD order. This effect will be discussed below.

	
	The eq. \eqref{eq variation volume} introduces the contact probability $p_{(i,j)}$ between two grain classes of the microstructure. The idea of Abbruzzese \textit{et al.} contact probability has been conserved in the equation form, however the spatial dimension of contact used is elevated to a volumic consideration as described in the following equation: 
	\begin{equation}
		p_{(i,j)} = \frac{N_j R_j^3}{\sum_{k=1}^{n_i}{N_k R_k^3}},
		\label{contact_probability_eq}
	\end{equation}
with $n_i$ the number of grain classes with an incomplete neighborhood when the $i\textsuperscript{th}$ grain class neighborhood is constructed.

	

	
	As presented previously, this model is described in 3D, so quantities exchanged during GB migration are volumic. The volume variation seen by a grain is the sum of all signed volume variations with respect to its neighbors given by eq. \eqref{eq variation volume} such as: 
	\begin{equation}
		\Delta V_i = \sum_{j=1}^{\eta_i} dV_{(i,j)}
	\end{equation}
	with $\eta_i$ the number of neighbors of the grain class $i$. As $dV_{(j,i)}=-dV_{(i,j)}$ owing to the imposed bijectivity, the volume conservation of the global system is ensured.

	\section{Input data for mean-field modeling}

	\subsection{Material-dependent model parameters acquisition}
	
	Experimental data and material parameters identification are necessary to calibrate a mean-field model for a given material and a given temperature range. This section will detail the material and experimental data used for model parameters identification.
	
	\subsubsection{Experimental data}
	A single-phase austenitic stainless steel (316L) was selected for this study. The identification procedure of the reduced mobility ($M_{GB}\gamma_{GB}$ product) presented in this work relies on a minimal campaign of nine thermal treatments \cite{alvarado_dissolution_2021,flipon_simplified_2022}. Fifteen are provided here and their conditions are detailed in table \ref{tab:minimum_doe_gg}. Longer annealing times from 3 to 5 hours have been realized to validate the identification procedure at long durations. These heat treatments have been performed using a Carbolite furnace. A thermocouple was placed in the furnace near to the samples to control and record the temperature evolution.

	\begin{table}[h!]
		\centering
		
		\begin{tabular}{|c|c|c|c|}
			\hline 
			\SI{1000}{\celsius} & \SI{1050}{\celsius} & \SI{1100}{\celsius} \\
			\hline 
            \arrayrulecolor{orange}
			\SI{30}{\min} & \SI{30}{\min} & \SI{30}{\min} \\ 
			\hline 
            \arrayrulecolor{red}
			\SI{1}{\hour} & \SI{1}{\hour} & \SI{1}{\hour} \\
			\hline 
			\SI{2}{\hour} & \SI{2}{\hour} & \SI{2}{\hour} \\
			\hline 
            \arrayrulecolor{green}
			\SI{3}{\hour} & \SI{3}{\hour} & \SI{3}{\hour} \\
			\hline 
			\SI{5}{\hour} & \SI{5}{\hour} & \SI{5}{\hour} \\
			\hline 
		\end{tabular}
		\caption{Heat treatments campaign conditions, with orange contoured conditions used as qualibration values, red conditions are used for both calibration and validation and green ones are only used for validation. }
		\label{tab:minimum_doe_gg}
	\end{table}
	
\begin{figure}[!h]
		\centering
		\begin{subfigure}[b]{0.49\textwidth}
			\includegraphics[width=\textwidth]{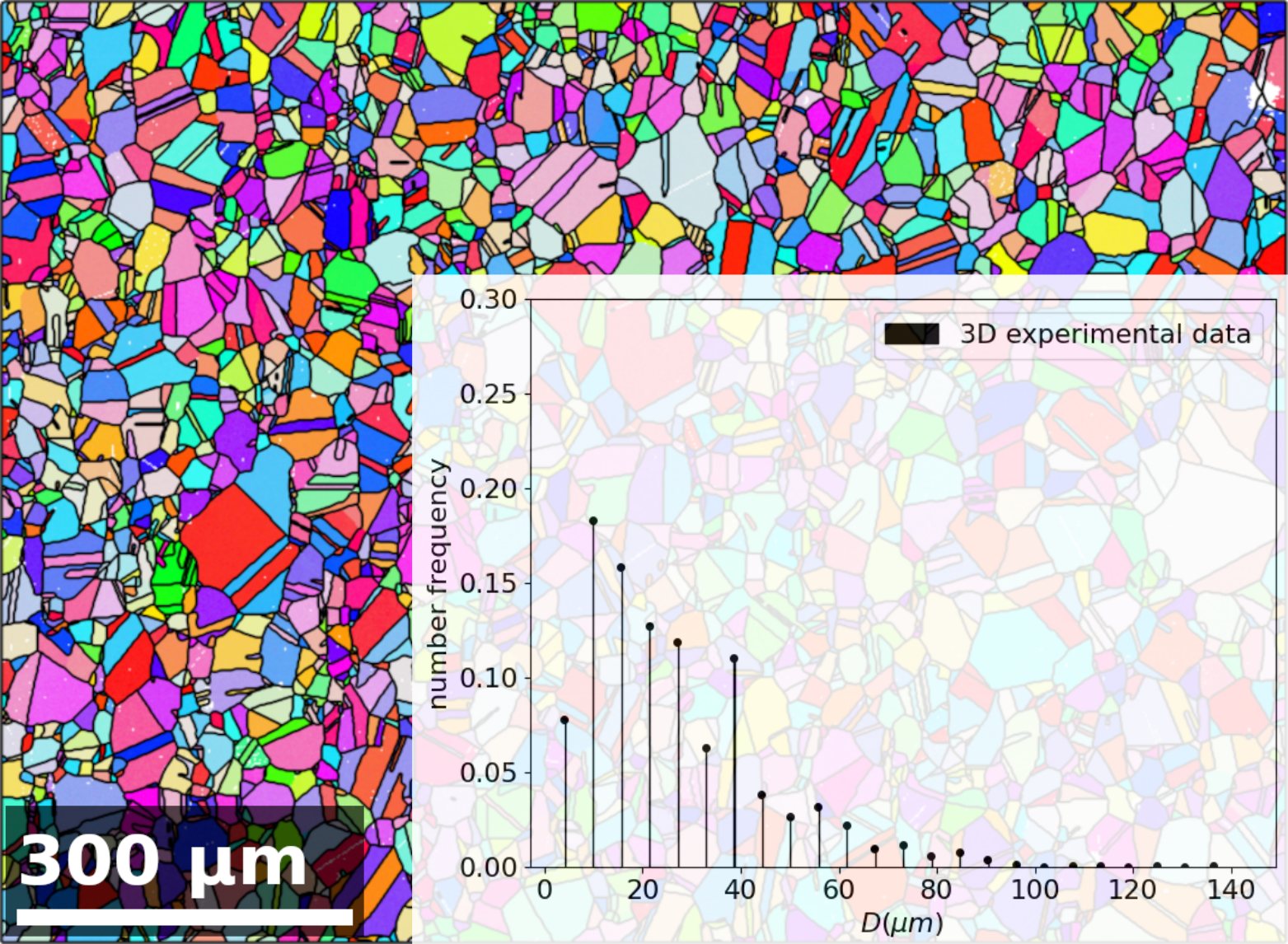}
			\caption{\label{}Initial microstructure.}
			\label{micro_ini}
		\end{subfigure}
		\begin{subfigure}[b]{0.49\textwidth}
			\includegraphics[width=\textwidth]{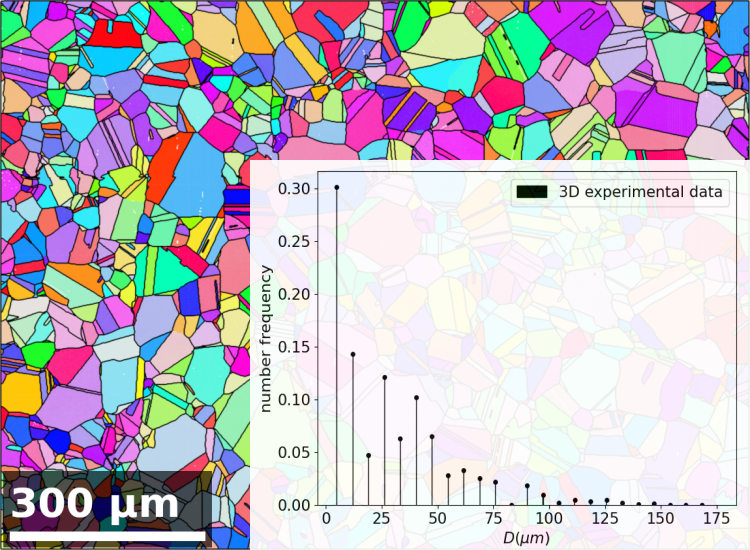}
			\caption{\label{}After a heat treatment of \SI{2}{\hour} at \SI{1000}{\celsius}.}
		\end{subfigure}
		\begin{subfigure}[b]{0.49\textwidth}
			\includegraphics[width=\textwidth]{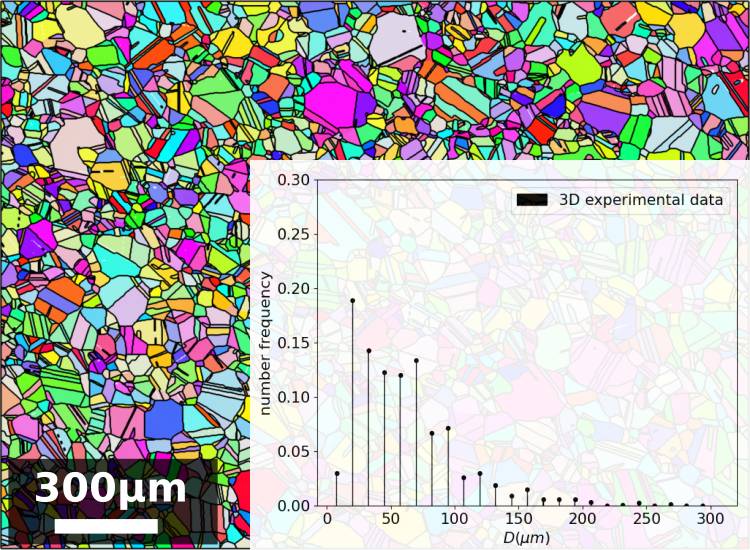}
			\caption{\label{H_1050_2h}After a heat treatment of \SI{2}{\hour} at \SI{1050}{\celsius}.}
		\end{subfigure}
		\begin{subfigure}[b]{0.49\textwidth}
			\includegraphics[width=\textwidth]{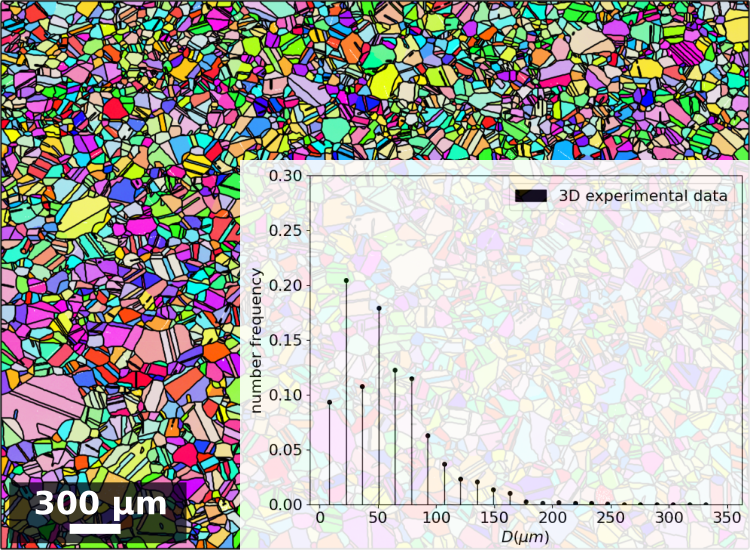}
			\caption{\label{M_1100_2h}After a heat treatment of \SI{2}{\hour} at \SI{1100}{\celsius}.}
		\end{subfigure}
		\caption{In background: EBSD IPF Z maps of 316L microstructures of (a) the as-received material, afer (b) 2h at 1000\degree C, (c)  2h at 1050\degree C and (d) 2h at 1100\degree C. Black lines denote all grain boundaries. In foreground for each image: the corresponding 3D-GSD after an inverse Saltykov transformation. }
		\label{fig:IPF hetero}
	\end{figure}

These samples have been prepared for electron back scattered diffraction (EBSD) analyses by cutting and selecting a centered observation area to avoid any effects of surface oxides on analyses. Classical first steps of polishing for stainless steel were realized using abrasive SiC papers, followed by polishing with a \SI{3}{\micro\meter} diamond suspension and final electropolishing for 25s at 10V with a solution of 10$\%$ perchloric acid in ethanol. EBSD analyses were performed with a Carl Zeiss Supra 40 field emission gun scanning electron microscope (FEGSEM) coupled with a Bruker Quantax EBSD detector and the Esprit 2.3 software. A voltage of 20kV and a \SI{120} {\micro\meter} aperture were used. The post-processing of EBSD data was achieved with the MTEX Matlab toolbox \cite{Bachmann2010}. The step and the cartography size have been targeted to get a representative number of grains in the observation area. Table \ref{tab:paramters_post_treatment} gathers the latter described parameters as well as the number of grains observed in each cartography. This number of grains is calculated without taking into account twin boundaries and using a misorientation of 15\degree to define a grain boundary. The minimal size to consider an entity as a grain is superior to 5 pixels. The process of entities removal under the set threshold and a grain boundaries smoothing were applied considering only indexed pixels. These data give also access to 2D GSD that will be transposed in 3D with the Saltykov algorithm (detailed in the following section) to be used as initial input distribution in the model but also as experimental data to compare with simulation results in section \ref{results}. Fig. \ref{fig:IPF hetero} displays the IPF Z maps of some of the post-treated experimental results as well as the related histograms in number frequency. 
 
	\begin{table}[h!]
		\small
		\begin{tabular}{|c|c|c|c|c|c|c|c|c|c|}
			\hline 
			\multicolumn{1}{|c|}{ \centering }& \multicolumn{3}{|c|}{ \centering $T=$1000\degree C } &\multicolumn{3}{|c|}{ \centering $T=$1050\degree C }&\multicolumn{3}{|c|}{ \centering $T=$1100\degree C } \\
            \hline
            \multicolumn{1}{|c|}{ $t$}& \multicolumn{1}{|c|}{ \centering $h$ (µm)} &\multicolumn{1}{|c|}{ \centering \tiny{$L_x\times L_y$ ($mm\times mm$)}}&\multicolumn{1}{|c|}{ \centering \#G }&
            \multicolumn{1}{|c|}{ \centering $h$ (µm)} &\multicolumn{1}{|c|}{ \centering \tiny{$L_x\times L_y$ ($mm\times mm$)} }&\multicolumn{1}{|c|}{ \centering \#G }&
           \multicolumn{1}{|c|}{ \centering $h$ (µm)} &\multicolumn{1}{|c|}{ \centering \tiny{$L_x\times L_y$ ($mm\times mm$)} }&\multicolumn{1}{|c|}{ \centering \#G }\\
           \hline 
           Initial&1.49&1.1$\times$0.85&980&1.49&1.1$\times$0.85&980&1.49&1.1$\times$0.85&980 \\
			\hline 
		      30min&2.5&2$\times$1.4&2654&1.13&1$\times$0.7&534&3.3&3.7$\times$2.8&3509 \\
			\hline 
            1h&2.5&2$\times$1.4&2078&3&3$\times$2.2&1964&3.3&3.7$\times$2.8&3590 \\
            \hline 
            2h&1.13&1$\times$0.7&456&3&3$\times$2.2&1154&3.77&3.7$\times$2.8&2208 \\
             \hline 
            3h&1.13&1$\times$0.7&468&1.13&1$\times$0.7&300&3.77&3.7$\times$2.8&2263 \\
             \hline 
            5h&1.13&1$\times$0.7&243&1.13&1$\times$0.7&133&3.77&3.7$\times$2.8&2304 \\
             \hline 
		\end{tabular}
		\caption{Table gathering, for the different isothermal treatment temperatures and for the different holding times, the post-processing details composed of the EBSD step size ($h$), the dimensions in each direction ($L_x\times L_y$) of analyzed areas of the sample, and the number of grains represented in the EBSD images considering all the grains without taking into account the twins boundaries (\#G).}
		\label{tab:paramters_post_treatment}
	\end{table}

\subsection{Use of Saltykov algorithm to obtain a 3D GSD} \label{salty}
 
\begin{figure}[!h]
		\centering
		\includegraphics[width=0.49\textwidth]{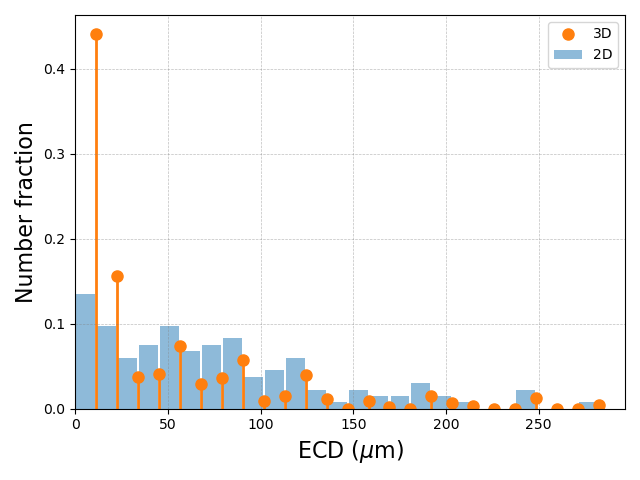}
		\caption{Inverse Saltykov method illustrated on the 2D GSD of a sample heat treated at \SI{1050}{\celsius} for \SI{5}{h}, the blue histogram corresponds to the 2D GSD and the orange discrete distribution corresponds to the obtained 3D results after the inverse Saltykov algorithm.}
        \label{fig:Saltky}
	\end{figure}

Experimentally acquired EBSD data correspond to 2D slices of 3D polycrystals. In order to be consistent with mean-field simulations results and to enable comparisons, a 2D to 3D conversion is performed. To this end, the Saltykov method \cite{Saltykov1958} is applied to experimental GSDs. Initially, the method had been developed to exhibit 2D section data from 3D granulometry data of spherical particles. The inverse Saltykov method gives the possibility to transform GSD data from a 2D histogram distribution to a 3D discrete distribution \cite{Saltykov1958}. Several assumptions are present in this method which are compatible with the topology of the considered initial 316L microstructure. The Saltykov method has indeed been proven to be efficient on a similar equiaxed polycrystal \cite{TUCKER2012554}. No assumption is required regarding the shape of the input distribution, so multi-modal distributions can also be submitted to such a method \cite{Lopez-Sanchez2016}. The methodology ensures that the average of an infinite number of 2D-cuts of a polycrystal respecting the obtained 3D-discrete distribution will converge towards the imposed 2D-histogram distribution. However, the quality of the methodology is, of course, also linked to the statistical representativity of the input 2D GSD. This procedure is illustrated in fig. \ref{fig:IPF hetero} to exhibit the 3D distribution evolution after the 2h thermal treatments at the different temperatures summarized in table \ref{tab:paramters_post_treatment}. Moreover, fig. \ref{fig:Saltky} illustrates for one particular microstructure ($T=$\SI{1050}{\celsius} for $t=$\SI{5}{h}) the comparison between the 2D-histogram distribution and the 3D obtained discrete distribution thanks to the inverse Saltykov transformation.\\

\subsubsection{GB mobility parameter identification} \label{fit_mob}
	
\paragraph{A first approximation by the use of classical B\&T law}

The first step of the GB mobility identification procedure is to find an initial approximation for the reduced mobility $(M_{GB} \gamma_{GB})$, in order to run a first mean-field computation. To this end, the historical form of the Burke and Turnbull law~\cite{BurkeTurnbull1952, Rollett2017} (eq. \eqref{parab_law_BT2} with $\tilde{\alpha}=1/2$ and $n=1$) is used to obtain a first approximated value of the reduced mobility for each considered temperature (see table \ref{tab:minimum_doe_gg}).
	
\begin{figure}[!h]
	\centering
           \begin{subfigure}[t]{0.49\textwidth}
			\includegraphics[width=\textwidth]{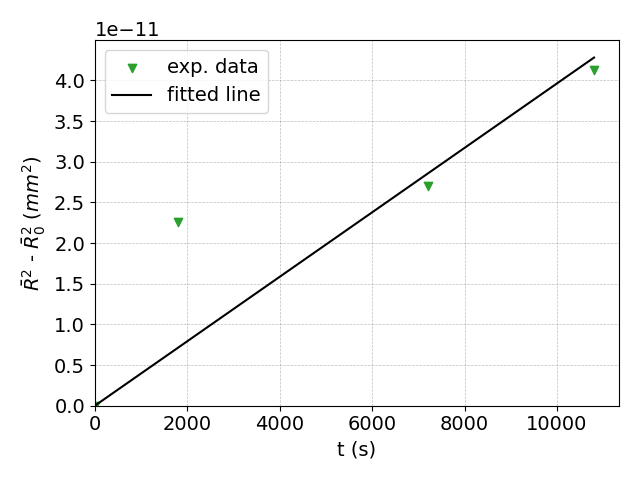}
			\caption{\label{1000_CBT}}
		\end{subfigure}
          \begin{subfigure}[t]{0.49\textwidth}
			\includegraphics[width=\textwidth]{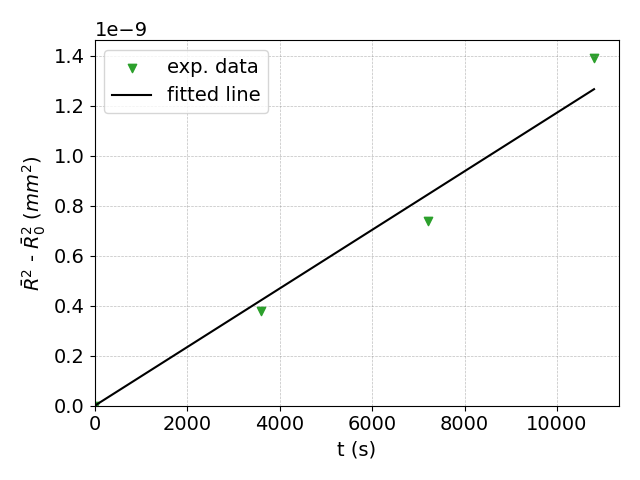}\caption{\label{1050_CBT}}
		\end{subfigure}
       \begin{subfigure}[b]{0.49\textwidth}
    		\includegraphics[width=\textwidth]{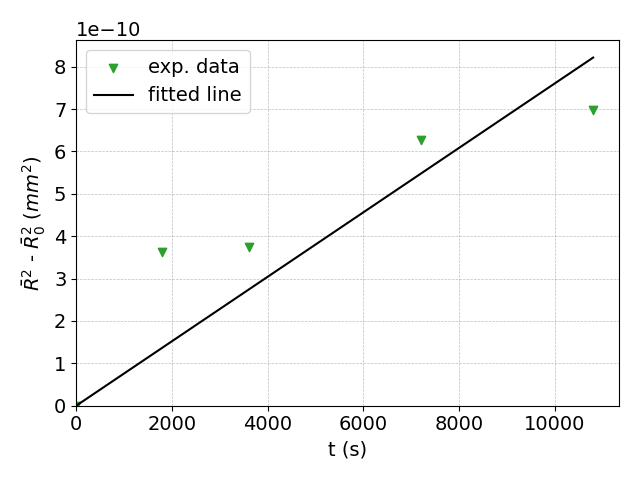}
    		\caption{\label{1100_CBT}}
    	\end{subfigure}
    \caption{ \label{B_T_figures}Use of the Burke \& Turnbull law to obtain the first value of $(M_{GB} \gamma_{GB})_{ini}$ for 316L at (a) \SI{1000}{\celsius},(b) \SI{1050}{\celsius} and (c) \SI{1100}{\celsius}.}
\end{figure}

For each heat treatment temperature, the B\&T law is plotted in order to obtain a linear dependence between $\bar{R}^2 - {\bar{R}_0}^2$ and the time $t$. Fig. \ref{B_T_figures} illustrates the methodology on sets of experimental points for the three studied temperatures where the best linear fit directly provides a first rough value for $(M_{GB} \gamma_{GB})$ called $(M_{GB}\gamma_{GB})_{ini}$.

\paragraph{Refined identification}
	
\begin{figure}[h!]
		\centering
		\includegraphics[width=0.60\textwidth]{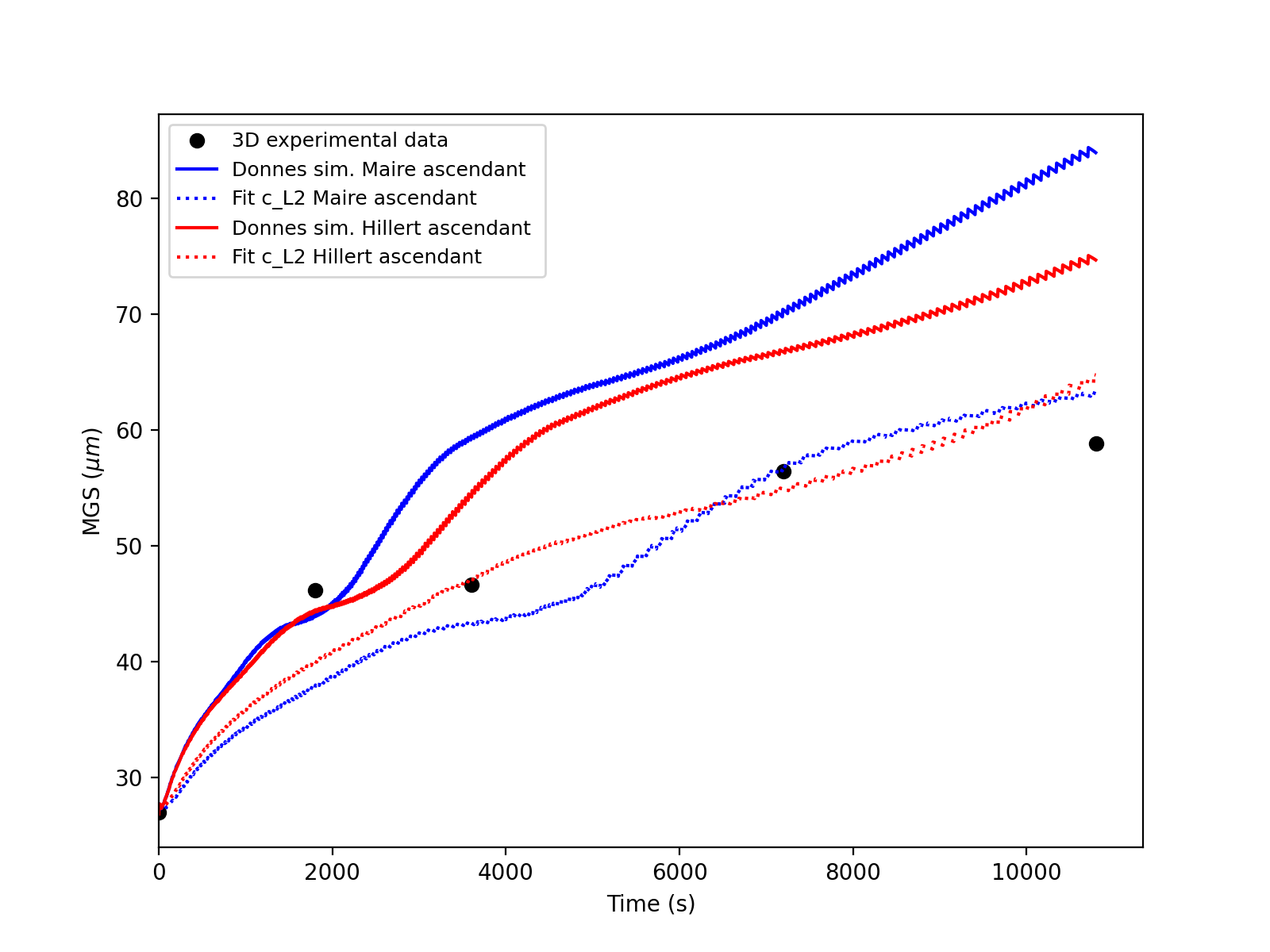}
		\caption{\label{mob_tool} Curve fitting of simulation points (obtained with $\left(M_{GB} \gamma_{GB}\right)_{ini}$) with respect to experimental points by minimizing \textit{L$^2$} error for Maire  \textit{et al.} and Hillert models at \SI{1050}{\celsius} for 316L.}
	\end{figure}

The values $(M_{GB}\gamma_{GB})_{ini}$ are then used in Hillert and Maire \textit{et al.} models for comparison with experimental points (solid blue and red lines respectively in fig. \ref{mob_tool}). These first simulation results are then used to perform an optimization of the reduced mobility by calculating the $L^2$ error on several points between the simulation and experimental results. A translation coefficient $c_{fit}$ is defined by a least square method to shift the simulated curve in order to improve the correlation with experimental data so that:
\begin{equation}
		\left(M_{GB} \gamma_{GB}\right)_{fit} = c_{fit} \times \left(M_{GB} \gamma_{GB}\right)_{ini}.
	\end{equation}

The cost function of the least square method is as followed: 

 \begin{equation}
  F(c)= \sum_{i=1}^{n} f_i^2(c_i),  
\end{equation}

where $n$ is the number of experimental points, $c_i=\frac{t_{sim_i}}{t_{exp_i}}$ the translation coefficient between the interpolated curve of the simulation data and the experimental points and $f_i(c_i)$ compute the $L^2$ errors between the simulated and experimental point for each value of $c_i$:

\begin{equation}
    f_i(c_i)=L^2_i(c_i)=100 \times \sqrt{\frac{\sum_{k=1}^n (\frac{t_{sim_k}}{c_i}-t_{exp_k})^2}{\sum_{k=1}^{n} t_{exp_k}^2}} \text{ with } \forall i \in \llbracket 1,n\rrbracket.
    \label{eq:L2_GBmob}
\end{equation}

	\paragraph{Reduced Mobility data}\label{Reducted mob}

	\begin{table}[h!]
		\centering
		\begin{tabular}{|c|c|c|c|}
			\hline 
			Temperature & \SI{1000}{\celsius} & \SI{1050}{\celsius} & \SI{1100}{\celsius} \\
			\hline 
			$M_{GB}\gamma_{GB}$(\SI{}{\meter\squared\per\second}) & 2.30e-15 & 1.08e-13  & 1.10e-13 \\
			\hline 
		\end{tabular}
		\caption{Table of the reduced mobility values $(M_{GB}\gamma_{GB})_{ini}$ for Maire \textit{et al.} model considering an ascending sorting order for the initial GSD for the investigated temperatures for 316L.}\label{tab mob}
	\end{table}

\paragraph{Model-dependence of Reduced Mobility}Since the intrinsic GB mobility of a material is hard to quantify experimentally, a common approach is to model the GB migration evolution (for instance using the $v = M_{GB}P$ equation) and consider the mobility $M_{GB}$ as a material-dependent model parameter. Depending on the definition used in the pressure term $P$, mobility will also be a model-dependent parameter.

Fig. \ref{mob_tool} solid lines illustrate the difference in response from Hillert and Maire \textit{et al.} models to an identical mobility value.
When the mobility is identified for each model, the respectively colored dashed lines are obtained for both models in the same figure. Reduced mobility values at the temperature of \SI{1100}{\celsius} are gathered in table \ref{tab_mob_models}. The GB migration equation in these models (cf. eq. \eqref{eq:Hillert}, \eqref{eq:dRij_Abb}, \eqref{eq:Pij}) is rather similar which explains that the identified reduced mobility are of the same order of magnitude.

     \begin{table}[h!]
        \centering
        \begin{tabular}{|c|c|c|c|}
            \hline 
            Model & Hillert & Abbruzzese & Maire \\
            \hline 
            $M_{GB}\gamma_{GB}$(\SI{}{\meter\squared\per\second}) & 1.08e-13 & 1.27e-13 & 1.10e-13 \\ 
            \hline 
        \end{tabular}
        \caption{Identified reduced mobility values for the different models at \SI{1100}{\celsius} for 316L.}
        \label{tab_mob_models}
    \end{table}

	\section{Results and discussion} \label{results}
	
 This section is dedicated to numerical parameters optimization and to the comparisons between results obtained using the different introduced mean-field models for 316L austenitic stainless steel in GG context. Different heat treatment conditions (temperature, duration) of table \ref{tab:minimum_doe_gg} are simulated and compared with experimental GSDs.

\subsection{Numerical parameters}

As described in section \ref{MFM}, the three mean-field models (Hillert, Abbruzzese  \textit{et al.} and Maire \textit{et al.}) are based on specific media or neighborhood. These latter may have an influence on several modeling parameters which requires a thorough study to either optimize their value or evaluate their impact on model predictions.

 \subsubsection{Convergence study concerning the number of grain classes introduced in the model} \label{conv}
An important common parameter to these models is the initial number of grain classes introduced at the start of a simulation. Statistical representativity is represented by two criteria: the minimum number of grain classes necessary for a GSD to be representative of an experimental microstructure and the representation necessity of the different grain populations existing in the material (detection of mono- or multimodal distributions). The convergence study performed here will focus on the first described criterion.
In GG, the number of grain classes drops in time due to capillarity effects, the representativity of the microstructure is therefore impacted. Maire \textit{et al.} neighborhood construction relies on a good statistical representativity, as it widens the choice of classes in the neighbors selection. 
A convergence study is performed on this parameter in order to determine the minimal initial number of grain classes necessary to obtain a reproducible final GSD.
A thermal treatment of one hour at \SI{1100}{\celsius} is used for this discussion and GSD results will be compared to a reference simulation for each model. More precisely, seven simulations from 25 to 2000  initial grain classes have been run and compared to a reference simulation where 5000 grain classes were considered. This reference is considered as representative of experimental values. Indeed, the number of grains in the sample area examined by EBSD analysis varies from 3500 to 130 grains with an average around 1400 grains for all studied conditions according to table \ref{tab:paramters_post_treatment}. The use of 5000 initial grains classes perfectly covers the 980 grains observed experimentally. Also, a final number of 1300 classes is obtained with this simulation after 5h. 

\begin{figure}[!h]
	\centering
           \begin{subfigure}[t]{0.49\textwidth}
			\includegraphics[width=\textwidth]{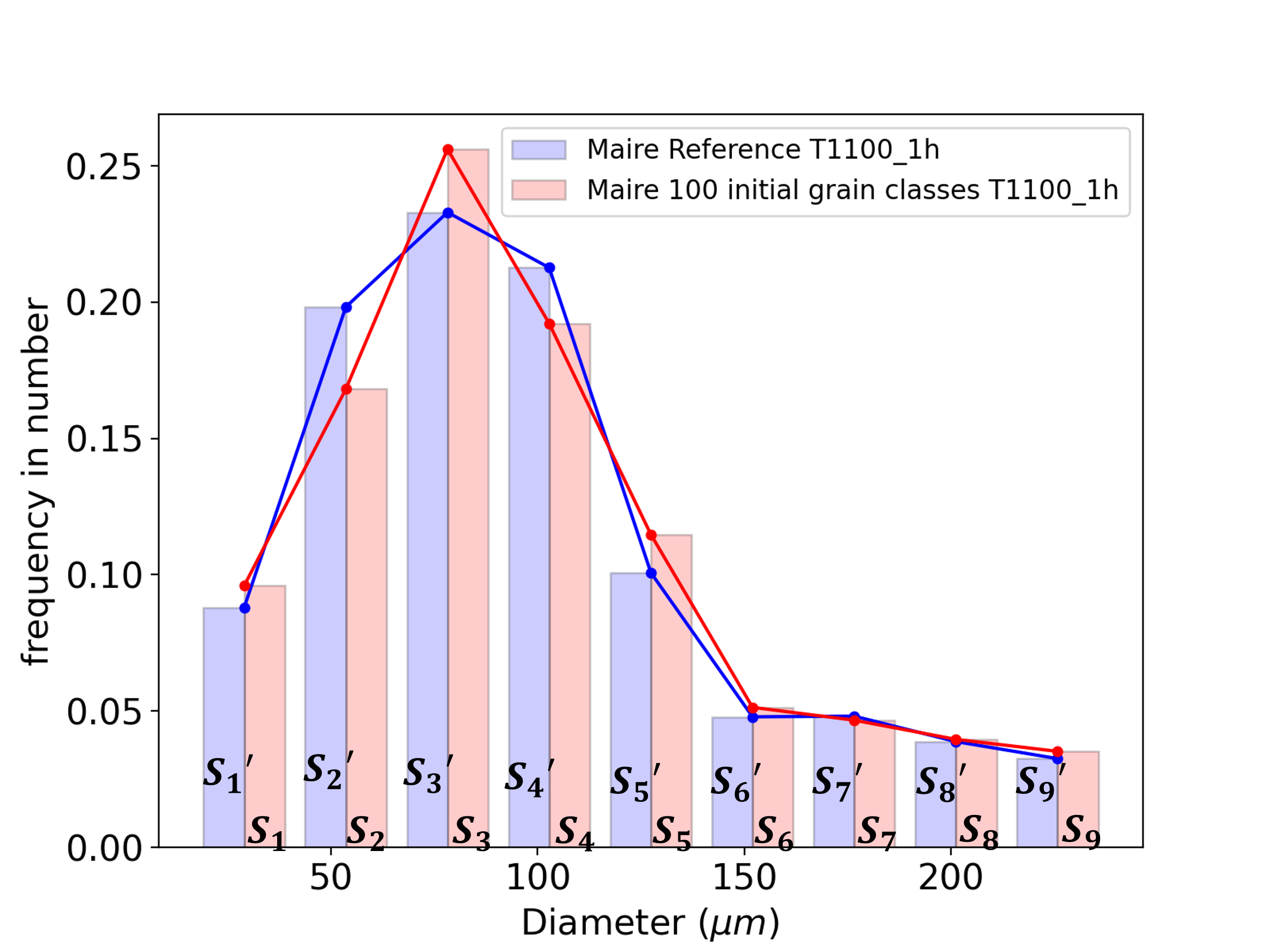}
			\caption{\label{L2_draw}}
		\end{subfigure}
          \begin{subfigure}[t]{0.49\textwidth}
			\includegraphics[width=\textwidth]{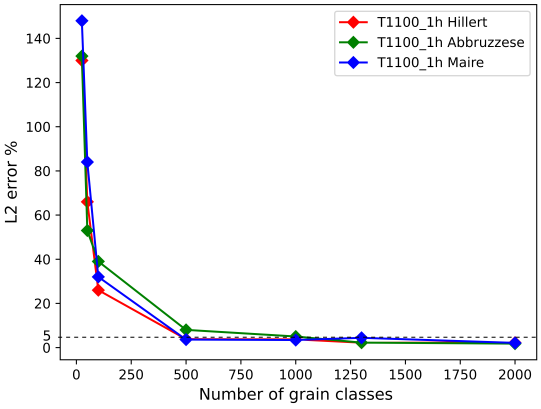}\caption{\label{convergence}}
		\end{subfigure}
       \begin{subfigure}[b]{0.49\textwidth}
    		\includegraphics[width=\textwidth]{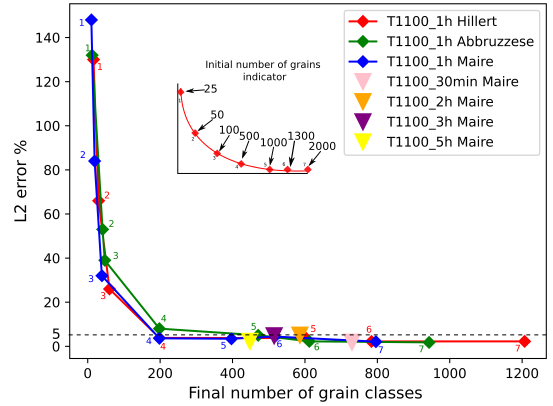}
    		\caption{\label{convergence_final_class}}
    	\end{subfigure}
    \caption{(a) Histogram description of the $L^2(1h)$ error method to analyse convergence of simulations at \SI{1100}{\celsius}, (b) convergence study of the initial grain classes number by the computing $L^2(1h)$ error at \SI{1100}{\celsius} for Hillert, Abbruzzese \textit{et al.} and Maire \textit{et al.} models on 316L and (c) same convergence study computing the $L^2(1h)$ error for the three models using the final number of grain classes as x-axis.}
\end{figure}

To be able to define at which value convergence is reached, eq. \eqref{eq:L2} gives the relative error, at time $t$ between the reference case and the tested GSDs:

\begin{equation}
		L^2(t) = 100 \times \sqrt{\frac{\sum_{i=1}^{n_{\text{bins}}}{(S_i - S_i')}^2}{\sum_{i=1}^{n_{\text{bins}}}{(S_i')}^2}},
		\label{eq:L2}
\end{equation} 
with $n_{\text{bins}}$ the number of bins in the histograms used for comparison. This number $n_{\text{bins}}$ is introduced to simplify the visual representation of histograms and have a meaningful comparison between simulations, a reduced number of histogram bins is then selected comparatively to the total number of grain classes. Typically, for the following histogram representations presented in this paper at exception of fig. \ref{L2_draw}, $n_{\text{bins}}$ is set to 25. The corresponding bin width is computed from this number. This allows to recreate a unique ECD vector of 25 visual grain classes equally distanced from each other to rightfully compare GSDs.

 In fig. \ref{L2_draw}, to introduce the $L^2$ comparison, a number of 9 bins is selected to simplify the visual explanation. In this figure $S_i$ (resp.  $S_i'$) corresponds, for the Maire \textit{et al.} model, to the $i\textsuperscript{th}$ bin grain class area of the GSD at $t=\SI{1}{\hour}$ (resp. of the model reference GSD).

Fig. \ref{convergence} plots the $L^2(\SI{1}{\hour})$ evolution for the different initial number of grain classes in each model at \SI{1100}{\celsius}. From 25 to 500 initial grain classes $L^2(\SI{1}{\hour})$ error decreases drastically from above 100$\%$ to 3$\%$. 25 initial grain classes give rise to a $L^2(\SI{1}{\hour})$ error superior to 100 $\%$ for all models, traducing therefore a degradation of the statistical representativity. A convergence threshold is set at 5$\%$, considering that convergence is reached below that error. From 500 to 2000 initial grain classes, the simulations are therefore converging considering the threshold. For comparisons performed in section \ref{results}, an initial number of 1000 grains classes is chosen in order to assure consistency and computational time efficiency.
This value is also in perfect accordance with the experimental initial number of grains of 980 as exposed in table \ref{tab:paramters_post_treatment}.
The statistical representativity depends also on a second parameter which corresponds to the number of grain classes present at the end of a numerical computation.

Fig. \ref{convergence_final_class} illustrates the same convergence study as in fig. \ref{convergence} but where the x-axis represents the number of final grain classes at the end of each simulation. This shows that a minimum number of 200 remaining grain classes is necessary to stay below a threshold of 5\% of $L^2(1h)$ error. As the model is working in grain classes, it can be considered that 200 classes can describe with good accordance the range of experimental number of grains from 130 to 3500 observed in EBSD maps for the different conditions. This figure also shows that for an identical initial number of grain classes Hillert model conserves a higher number of remaining grain classes and therefore a better statistical representativity. 

For forthcoming GSD comparisons, the statistical representativity for heat treatment from 2 to \SI{5}{\hour} at \SI{1100}{\celsius} with an initial number of 1000 grain classes was also verified thanks to the Maire \textit{et al.} model as illustrated in fig. \ref{convergence_final_class} with the colored triangle icons. Indeed, one can see for the three thermal treatments that the threshold for representativity previously defined is well respected as the $L^2(t)$ error (comparatively to the reference case at 5000 grains classes) remains below 5\% and that the remaining number of classes after annealing time is above 200.

\subsubsection{Different spatial dimensions considered to define the contact probability} \label{diff contact prob}

\paragraph{Description of the spatial dimensions}

In the original work of Abbruzzese \textit{et al.} \cite{Abbruzzese1992}, the 2D contact probability is computed based on the perimeter of the neighbor class (cf. section \ref{subsec:Abb_2D}). This formalism was extended to 3D by considering sphere surfaces instead of disk perimeters in Di Schino \textit{et al.} \cite{DiSchino2001}. The contact probability construction suggested by Maire \textit{et al.} is a generalization of the latter formalism. This section will focus on determining if the contact probability has a positive effect on refining the description of Maire \textit{et al.} model simulated GSDs. 
Inspired from Abbruzzesse \textit{et al.} work with eq. \eqref{contact_prob_2D_abb}, four types of contact probability can be derived from a probability in number to a volumic probability:  
	
	\begin{equation}
		p_{(i,j)}^{m} = \frac{N_j R_{j}^{m}}{\sum_{k=1}^{n_i}{N_k R_{k}^{m}}}\quad \text{with } m\in\llbracket 0,3\rrbracket\label{p_j}, 
	\end{equation}
	where $N_j$ is the number of grains belonging to the grain class $j$ and $n_i$ the number of grain classes with an incomplete neighborhood when $p_{(i,j)}^{m}$ is computed for the grain class $i$.

 	\begin{figure}[!h]
    	\centering
    	\includegraphics[width=0.49\textwidth]{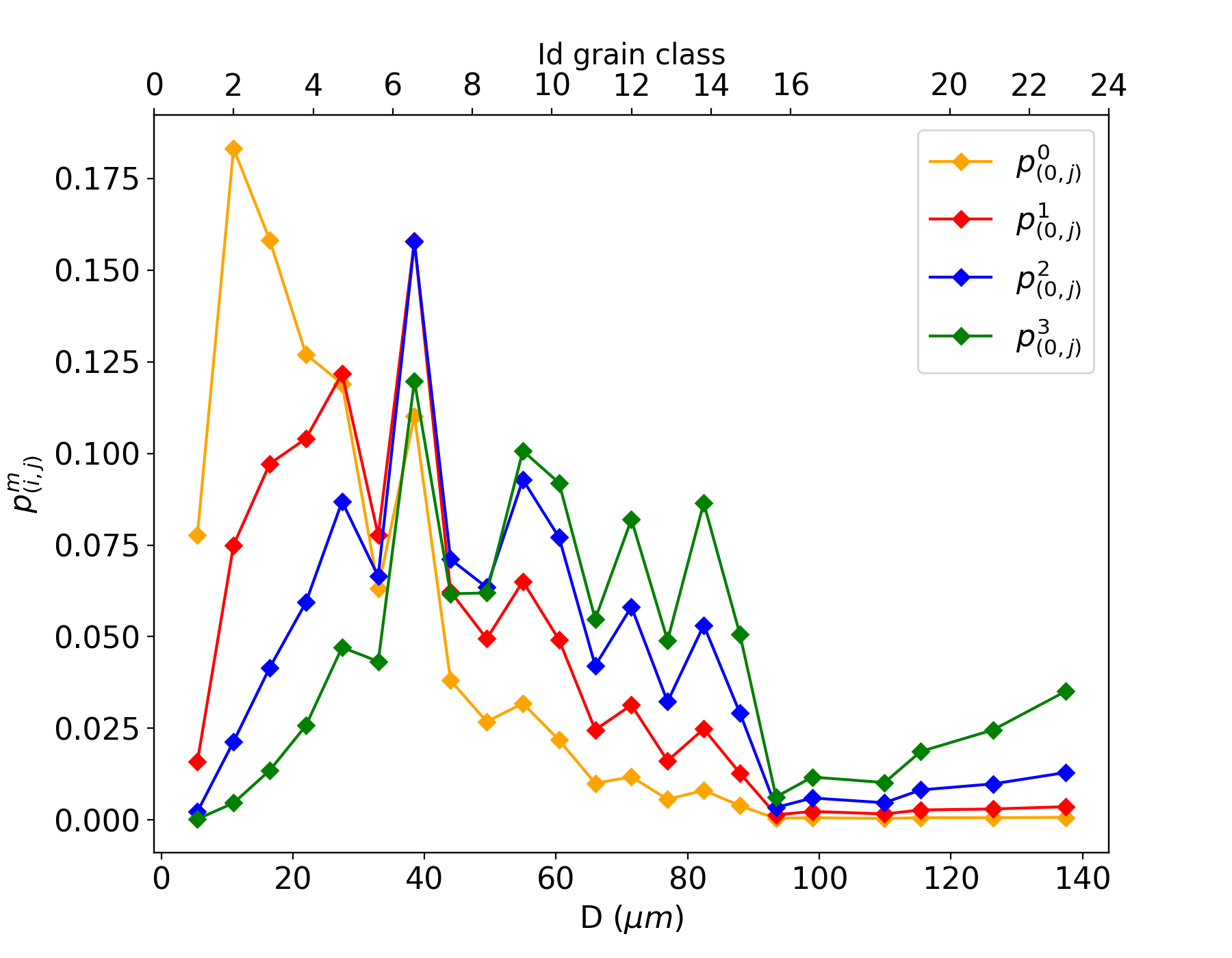}
        \caption{\label{fig:dep_en_R_CP}Description of the different contact probabilities $p_{(i,j)}^{m}$ with neighbor classes of the first class ($i=0$) of the microstructure at t=\SI{0}{s}.}
	\end{figure}

The four contact probability definitions described by eq. \eqref{p_j} are defined for the neighborhood of the first grain class of the microstructure at the initial time t=\SI{0}{s}, as plotted in fig. \ref{fig:dep_en_R_CP}. The probabilities are used in the context of Maire \textit{et al.} model detailed in section \ref{Maire_model} by modifying eq. \eqref{contact_probability_eq}. In eq. \eqref{p_j} with $m=0$, the contact probability $p_{(i,j)}^{0}$ is computed in terms of number with no influence of the grain size. In fig. \ref{fig:dep_en_R_CP}, the orange curve shows that the majority of the weight is given to smaller neighbor grains as only the frequency of occurrence in the microstructure is taken into account. The $p_{(i,j)}^{1}$ formulation uses the perimeter of the grain class to describe the contact probability. This latter is taking the highest probability values on the first one-third grain classes, which traduces that small grain classes are still more represented in the neighborhood of the first class in this description. $p_{(i,j)}^{2}$ formulation considers the surface of the neighbor grain class $j$ to build the contact probability. A shift toward the middle grain sizes is observed, leading to larger grains taking part of the neighborhood compared to $p_{(i,j)}^{0}$ and $p_{(i,j)}^{1}$ definitions. Finally, the description $p_{(i,j)}^{3}$ imposes a neighborhood based on the volume of the neighbor grain classes which gives more weights to larger grain classes as highlighted by the green curve in fig. \ref{fig:dep_en_R_CP}. Looking solely at neighborhood construction, none of these representations seems more justified than others. Depending on the selected representation, an emphasis on certain groups of grains is done as explained above.
	
\paragraph{Impact on the distribution results}
	
\begin{figure}[!h]
		\centering
		\begin{subfigure}[b]{0.49\textwidth}
			\includegraphics[width=\textwidth]{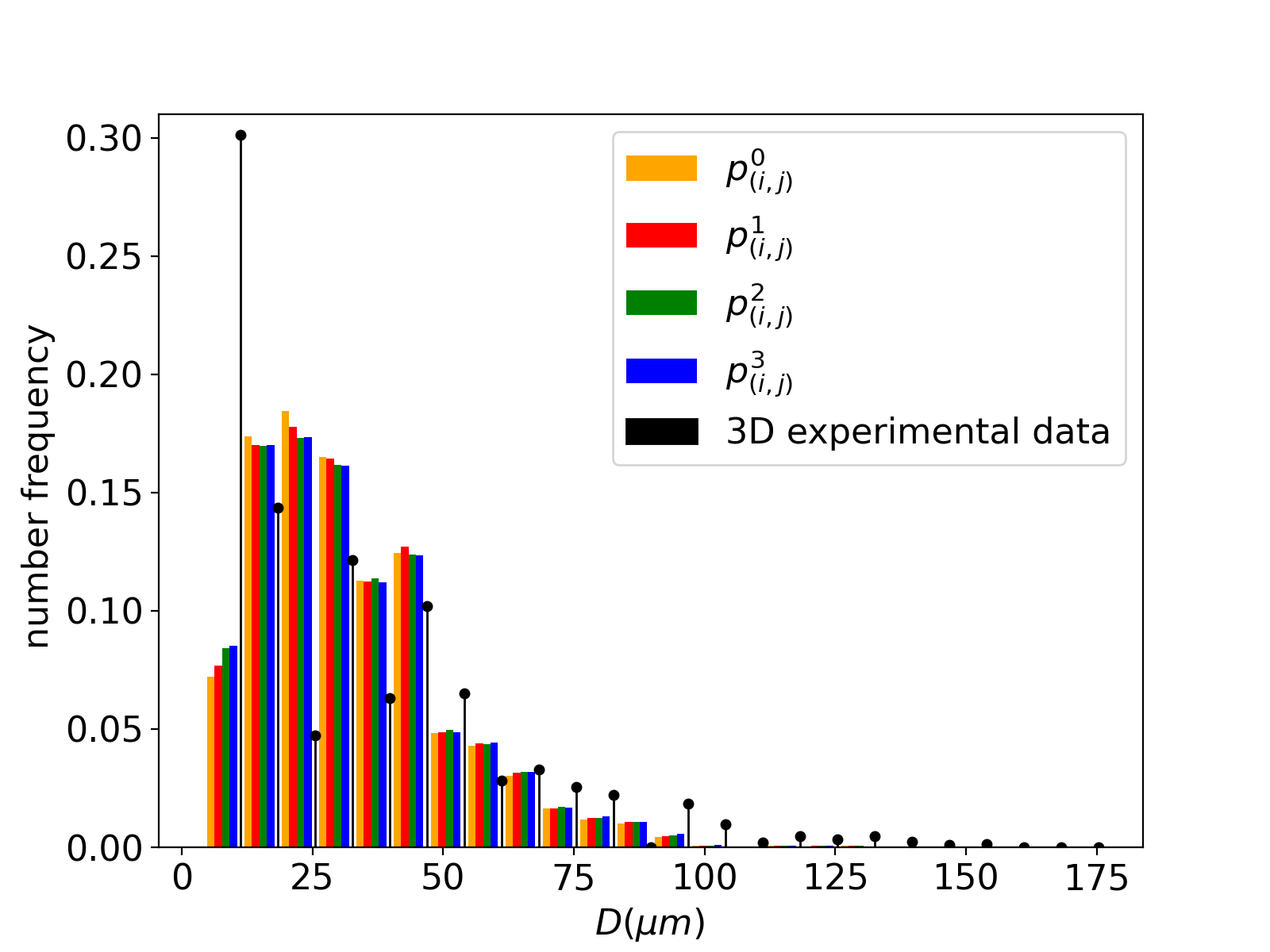}
			\caption{\label{dep_en_R_plot_1000_2h}\SI{1000}{\celsius}}
		\end{subfigure}
		\begin{subfigure}[b]{0.49\textwidth}
			\includegraphics[width=\textwidth]{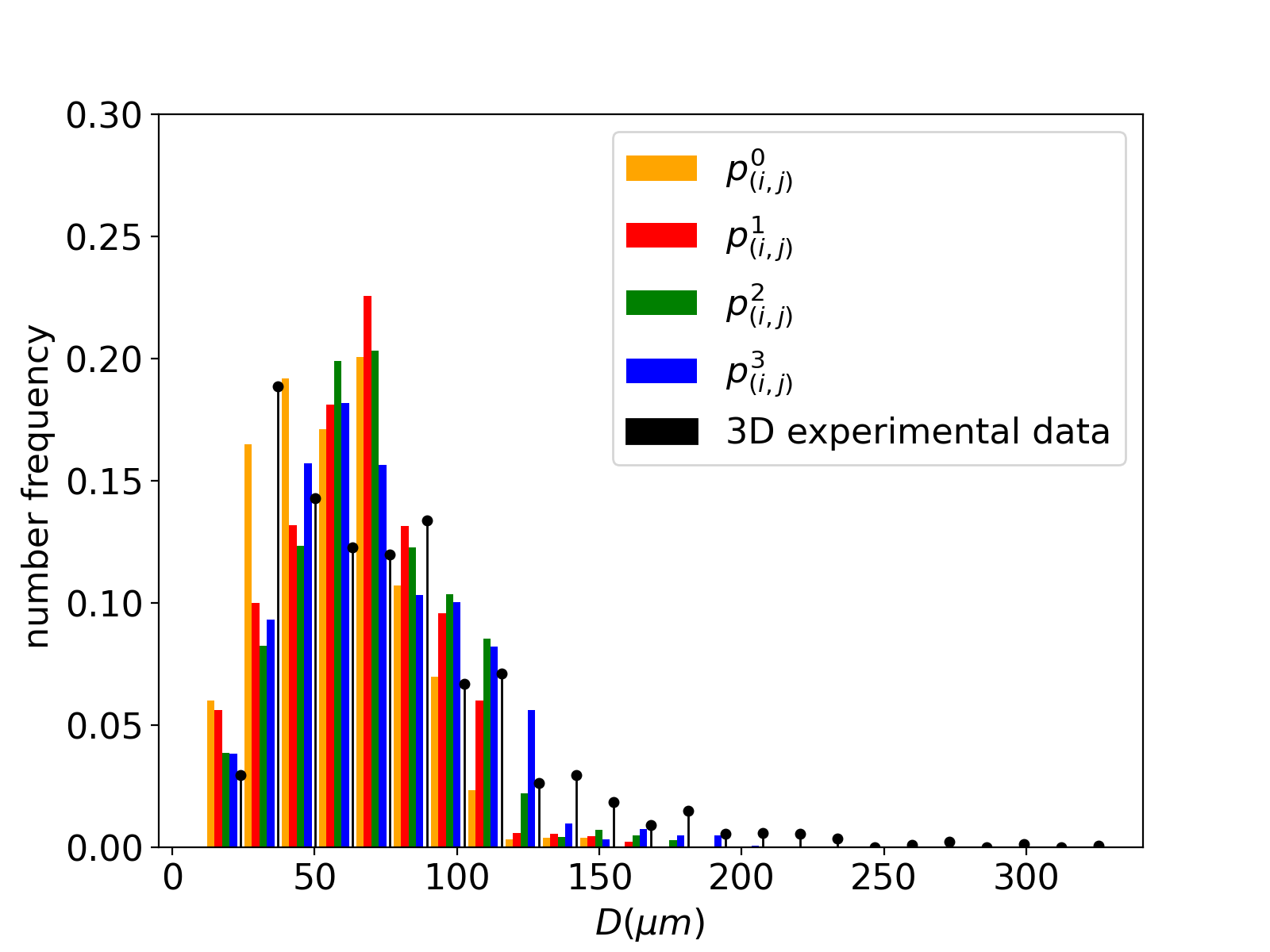}
            \caption{\label{dep_en_R_plot_1050_2h}\SI{1050}{\celsius}}
		\end{subfigure}
		\begin{subfigure}[b]{0.49\textwidth}
			\includegraphics[width=\textwidth]{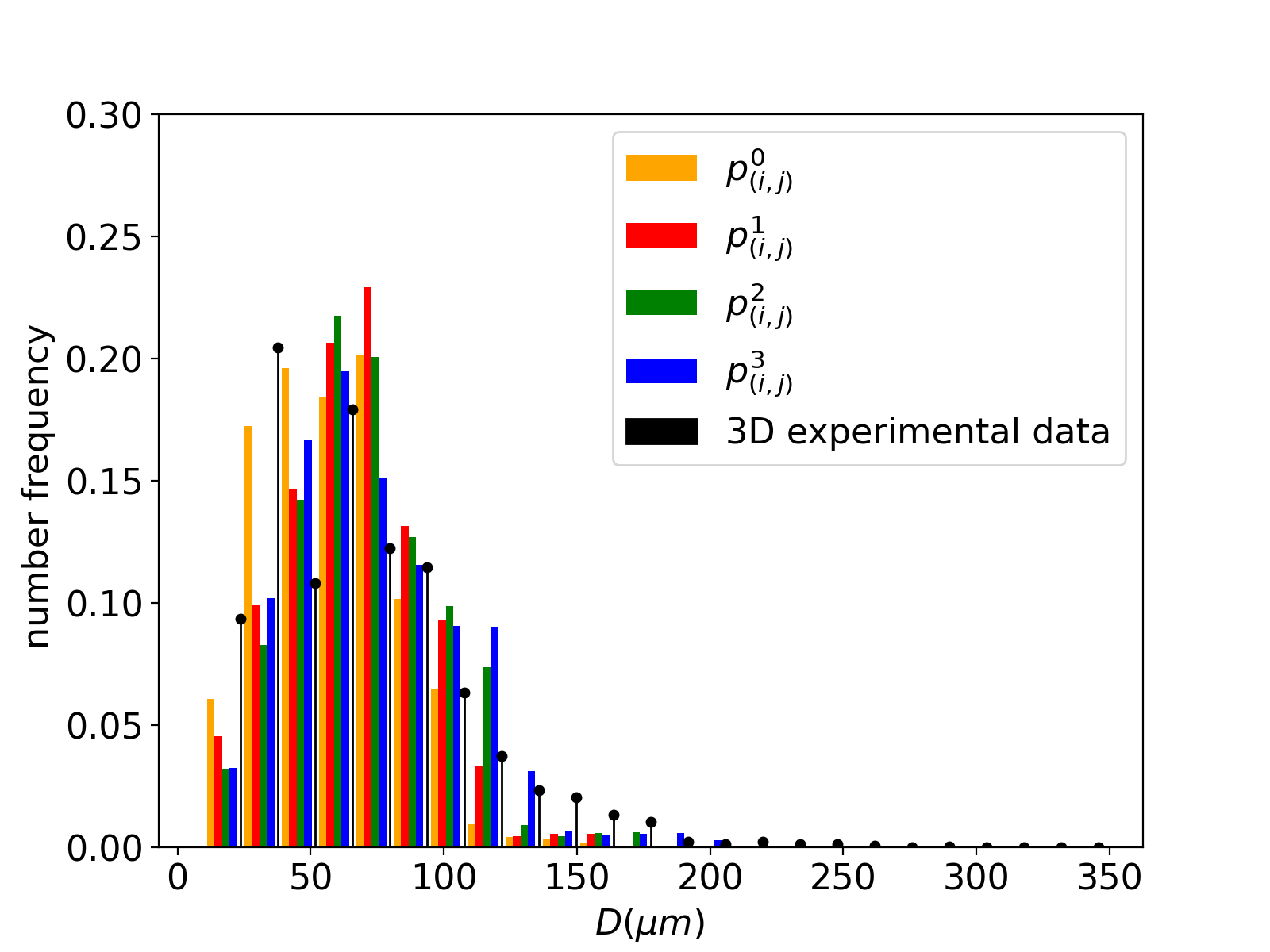}
			\caption{\label{dep_en_R_plot_1100_2h}\SI{1100}{\celsius}}
			
		\end{subfigure}
		
		\caption{Comparison of the impact on the distribution of the $p_{(i,j)}^{m}$ with $m \in \llbracket{0,3}\rrbracket $ for an annealing of \SI{2}{\hour} at different temperatures for 316L.}\label{fig:remplissage_voisin}
	\end{figure}

To select one of the above contact probability definitions for the study, grain size distribution results will be compared using a GG test case. For the three investigated temperatures, 1000, 1050 and 1100\degree C an annealing of two hours is simulated with the four contact probabilities. The GSD results are compared to the data converted into a 3D GSD in fig. \ref{fig:remplissage_voisin} using the Saltykov method described in section \ref{salty}. The experimental data are represented by the discrete black histogram. At all three temperatures, $p_{(i,j)}^{3}$ seems to provide a better fit of the tail of the distribution. This volume-based description brought by eq. \eqref{p_j} with $m=3$ accentuates the topological effect of the neighborhood construction by giving more weight to bigger grains. If a homogeneous microstructure is composed solely of small grains, the contact probability description will give similar contact probabilities to these grains with a comparable volume. However, if a heterogeneous microstructure with large and small grains is considered, the volumic probability will bring a topological aspect by giving more weight to larger grains. Indeed, if bigger grains have a better representation in the neighborhood of a grain class, then the $dV$ exchange achieved with the GB migration eq. \eqref{eq variation volume} for these grain classes is increased. Therefore, they have statistically more chances to grow and be represented in the distribution in further time steps.

\subsubsection{Impact of the grain classes order in the neighborhood construction}\label{homo_descending}

The complexity of the neighborhood construction proposed by Maire \textit{et al.} is influenced by the selecting order of the grain classes in the GSD. As mentioned in section \ref{Maire_model}, in Maire \textit{et al.} original work \cite{Maire2018} the ascending order has been arbitrarily selected for the GSD. For Hillert and Abbruzzese \textit{et al.} models, the MGS evolution with respect to time shows no changes no matter in which order the grain classes are selected, as no topology is involved in the computation of their surrounding media.  

To study the effect of the selecting order of the grain classes in the neighborhood construction on the results, two other types of selecting order are considered: GSD is either selected by decreasing grain sizes or randomly.

\begin{figure}[!h]
		\centering
            \begin{subfigure}[b]{0.49\textwidth}
			\includegraphics[width=\textwidth]{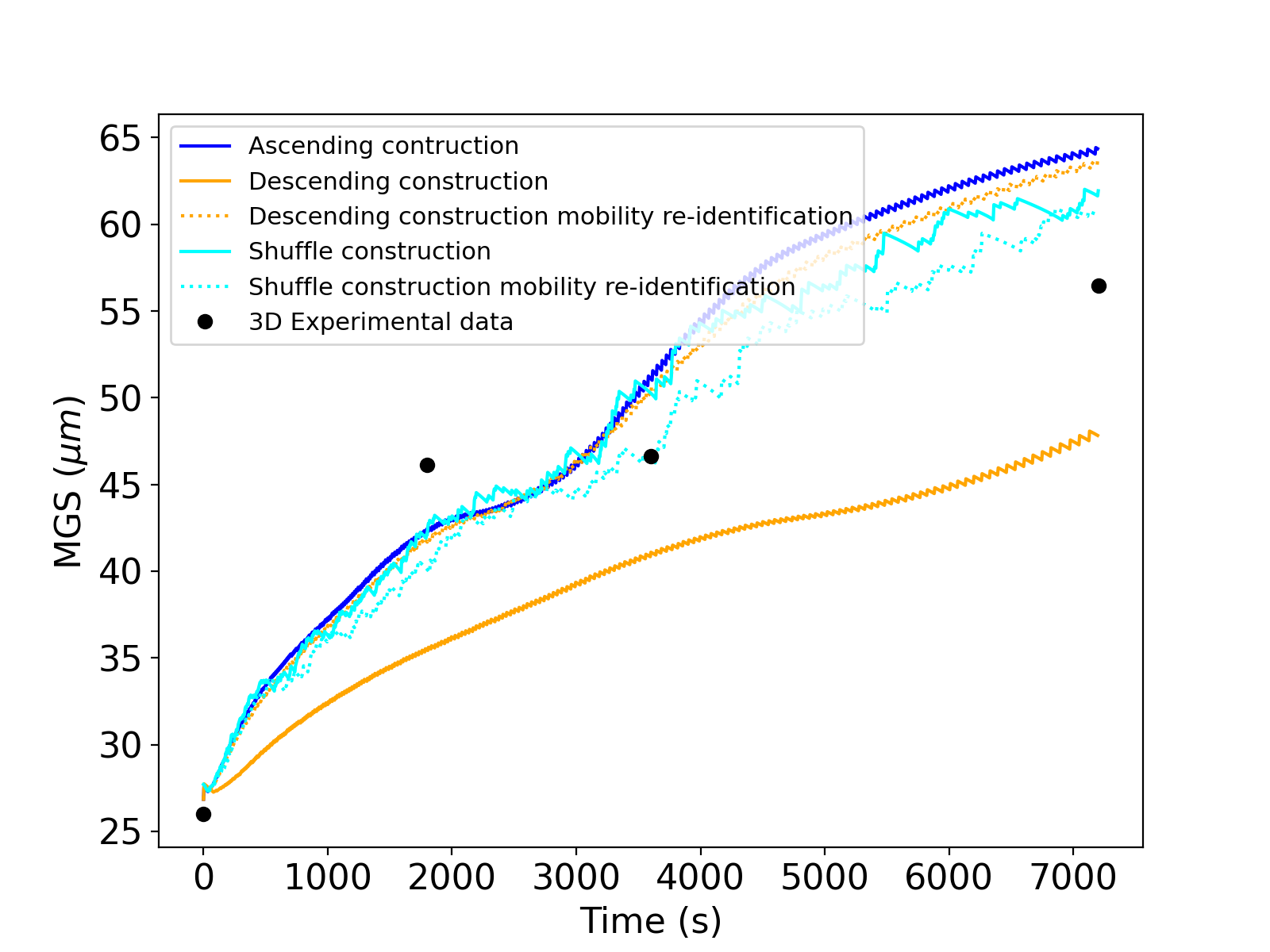}
			\caption{\label{GSD_comp_order} }
		\end{subfigure}

      \begin{subfigure}[b]{0.49\textwidth}
			\includegraphics[width=\textwidth]{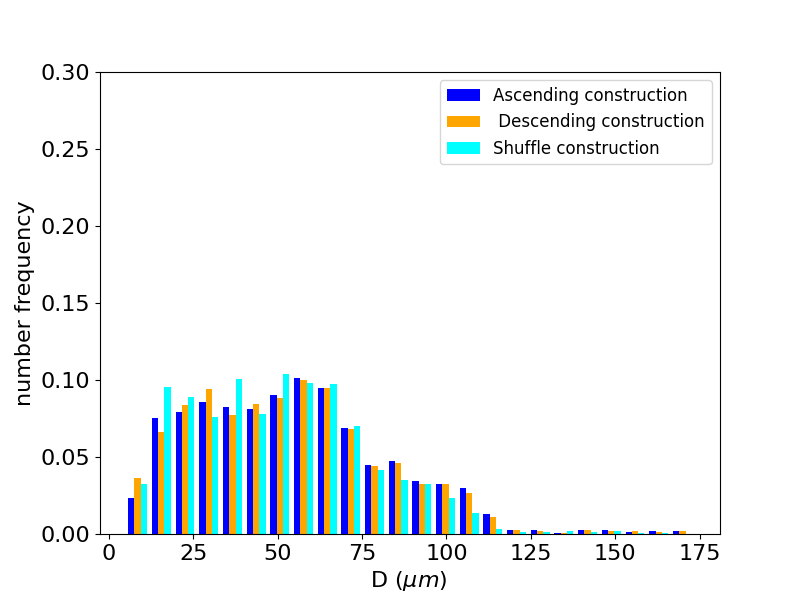}
			\caption{\label{GSD_ac_reid_nb} }
		\end{subfigure}
    \begin{subfigure}[b]{0.49\textwidth}
			\includegraphics[width=\textwidth]{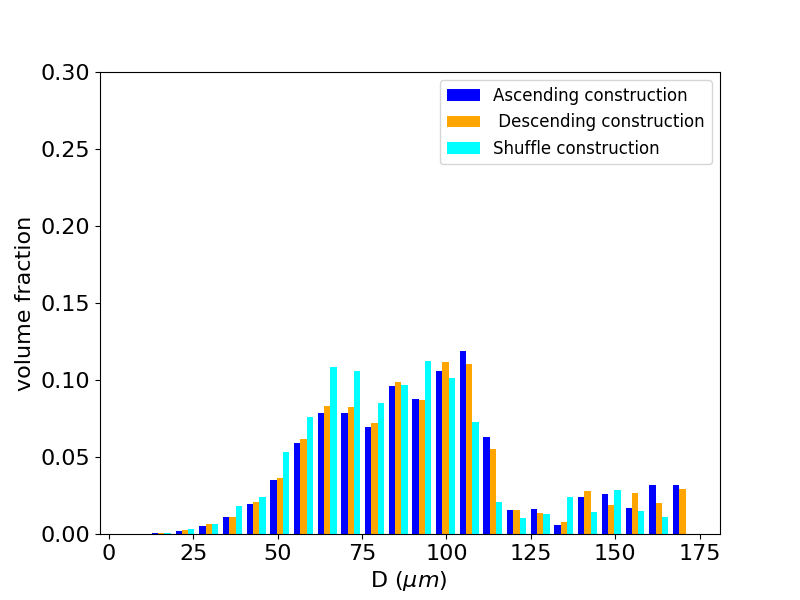}
			\caption{\label{GSD_ac_reid_vol} }
		\end{subfigure}
  \caption{Comparing different selecting orders for the neighborhood construction using the Maire \textit{et al.} model. Test case selected here is a \SI{1}{\hour} annealing at \SI{1100}{\celsius}. (a) MGS evolution with respect to time, GSD at the end of the heat treatment (t=\SI{1}{\hour}) considering (b) number frequency and (c) volume fraction.}
\end{figure}

\begin{figure}[!h]
	\centering
    \begin{subfigure}[b]{0.4\textwidth}
		\includegraphics[width=\textwidth]{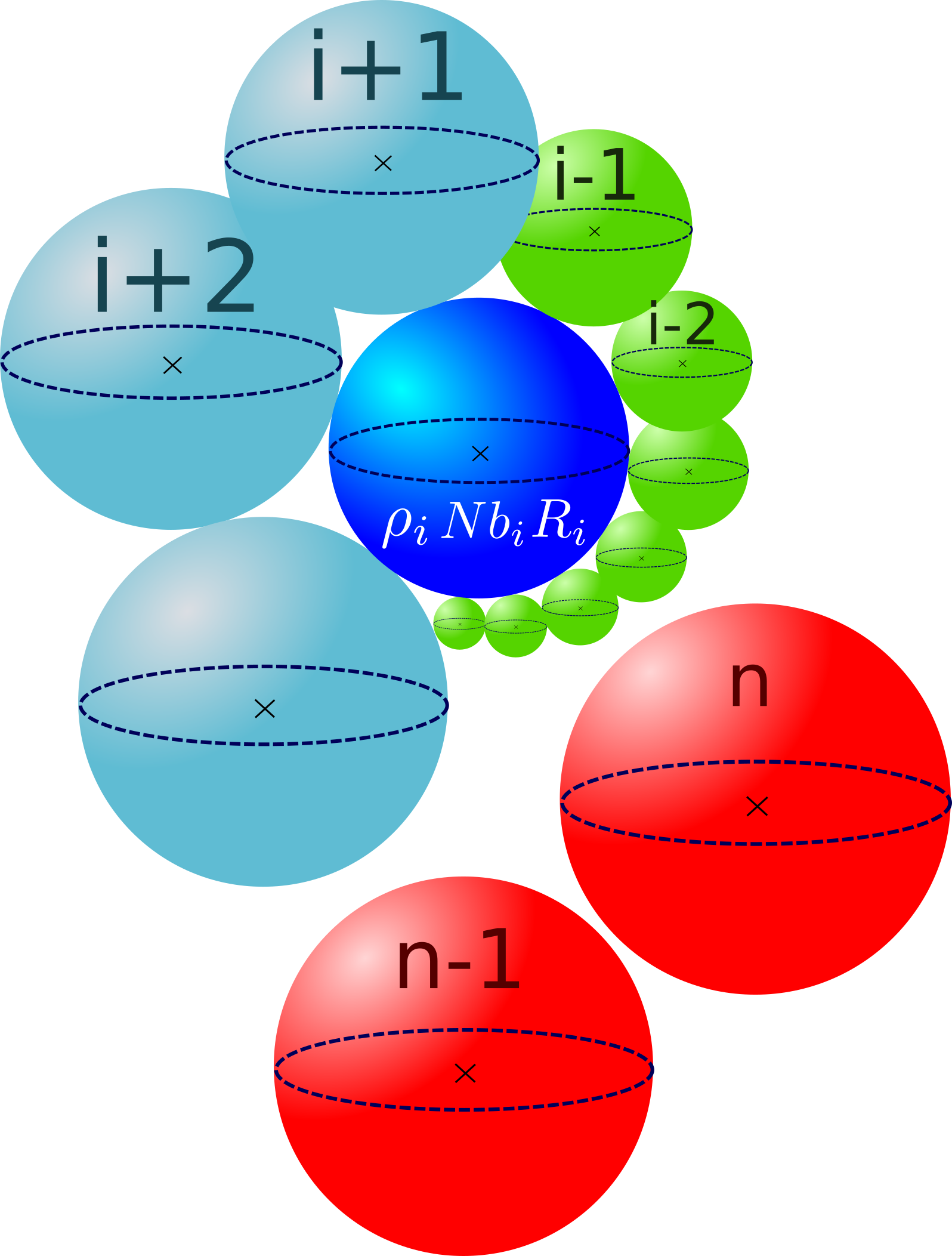}
		\caption{\label{Neighbor_construction_ascending}}
    \end{subfigure}
    \begin{subfigure}[b]{0.4\textwidth}
		\includegraphics[width=\textwidth]{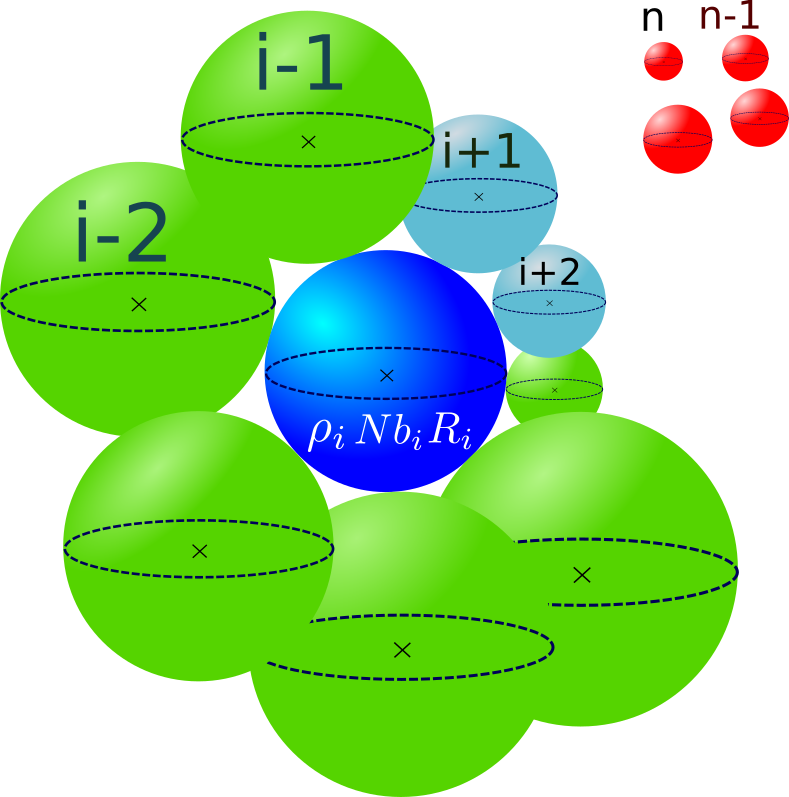}
		\caption{\label{Neighbor_construction_descending}}
    \end{subfigure}
    \caption{Specific neighborhood construction considering (a) the ascending and (b) the descending selecting order in the GSD. \label{contruction_oder}}
\end{figure}

In fig. \ref{GSD_comp_order}, the MGS evolution is strongly affected by the selection in descending sorting order for construction of the neigborhood as described by the orange solid line. The shuffle order of construction has a smaller impact on the MGS kinetic compared to the latter one. The reduced mobility values need to be re-identified to retrieve a good fit with the experimental data. The values for these specific selecting orders of construction in Maire \textit{et al.} model are gathered in table \ref{tab_mob_order} and the associated MGS evolutions are represented by the dashed lines in fig. \ref{GSD_comp_order}. The corresponding GSDs are presented in terms of number frequency and volume fraction in fig. \ref{GSD_ac_reid_nb} and \ref{GSD_ac_reid_vol}.
When the reduced mobility is re-identified, predicted GSDs remain close to each other independently of the selected sorting order of construction. Ascending and descending construction orders tend to predict longer distribution tails than the shuffle order. However, if the mobility is not re-identified, the strong impact for MGS prediction can be explained by the difference in neighborhood construction for the same current grain class. The microstructure representations of these construction orders are schematically presented in fig. \ref{Neighbor_construction_ascending} and fig. \ref{Neighbor_construction_descending}, describing two different construction patterns. In ascending sorting order, bigger grains of the microstructure are red grain classes, \textit{i.e.} the ones that will not take part in the neighborhood of the current class $i$. On the contrary, in the descending sorting order construction, red grain classes are the smaller ones. This will modify the GB migration  volume exchanges between neighboring grains during GB migration.

 For the forthcoming comparisons, the original ascending GSD sorting order will be conserved, as in this case the reduced mobility value has a similar order of magnitude to the one of the other models.
 
\begin{table}[!h]
		\centering
		\begin{tabular}{|c|c|c|c|}
			\hline 
			Sorting order & Ascending & Descending & Shuffle \\
			\hline 
			$M_{GB}\gamma_{GB}$(\SI{}{\meter\squared\per\second}) & 2.19e-13 & 5.00e-13 & 2.30e-13 \\
			\hline 
		\end{tabular}
		\caption{Identified reduced mobility values for Maire  \textit{et al.} model for different selecting order at \SI{1100}{\celsius} for 316L.}\label{tab_mob_order}
	\end{table}

\subsection{Comparison of mean-field models using different initial microstructures}

In this section, initial mono- and bimodal distributions will be used to evaluate the impact of heterogeneities in the initial GSD. The monomodal test case use the initial experimental GSD. Thermal treatment simulations will be compared to experimental data. However, bimodal comparisons are confronted only between themselves as no experimental data were available for this case. In both analyses, Maire  \textit{et al.} model is employed with a volumic contact probability as detailed in section \ref{diff contact prob} and with an initial number of grain classes of 1000 as deduced from section \ref{conv}. An ascending sorting order for the input distribution is considered as originally used in previous Maire \textit{et al.} work \cite{Maire2018}.

\subsubsection{Comparison of mean-field models with a monomodal initial microstructure}

\begin{figure}[!h]
		\centering
		\begin{subfigure}[b]{0.35\textwidth}
			\includegraphics[width=\textwidth]{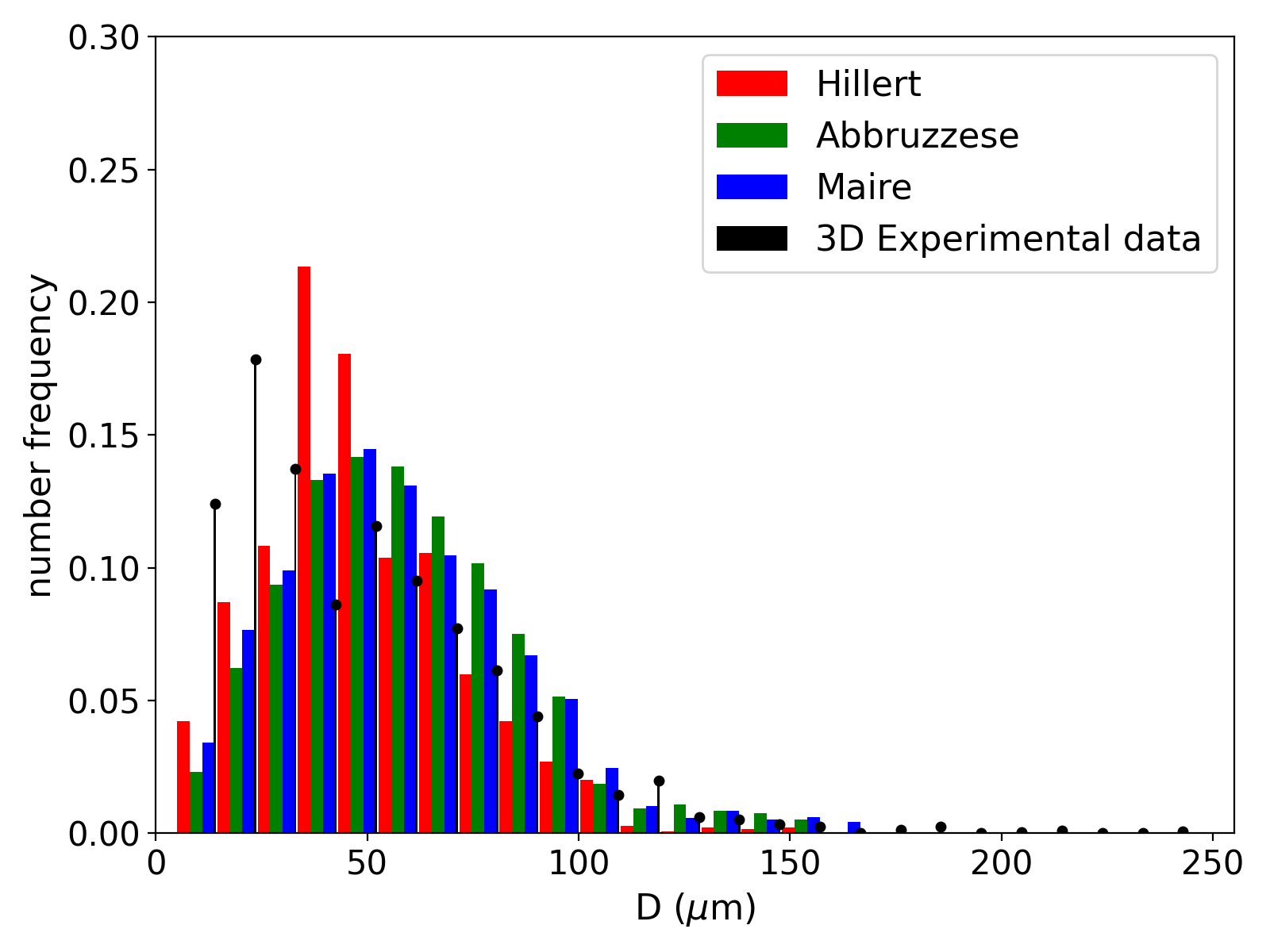}
			\caption{\label{1100_surf_1h_diam} 1h at \SI{1100}{\celsius}.}
		\end{subfigure}
        \begin{subfigure}[b]{0.35\textwidth}
    			\includegraphics[width=\textwidth]{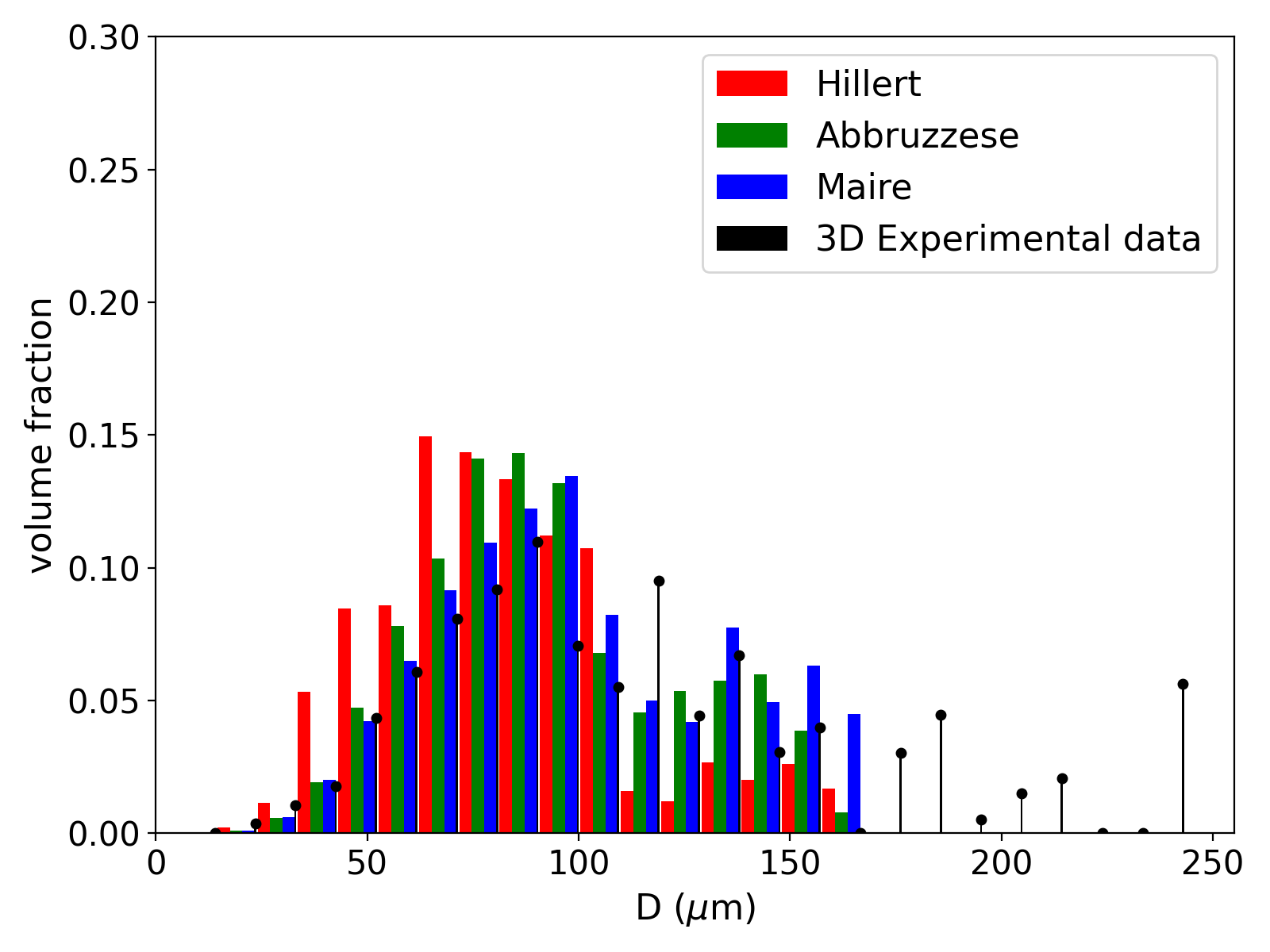}
    			\caption{\label{1100_surf_1h_diam_surf} 1h at \SI{1100}{\celsius}.}
    		\end{subfigure}
      
    		\begin{subfigure}[b]{0.35\textwidth}
    			\includegraphics[width=\textwidth]{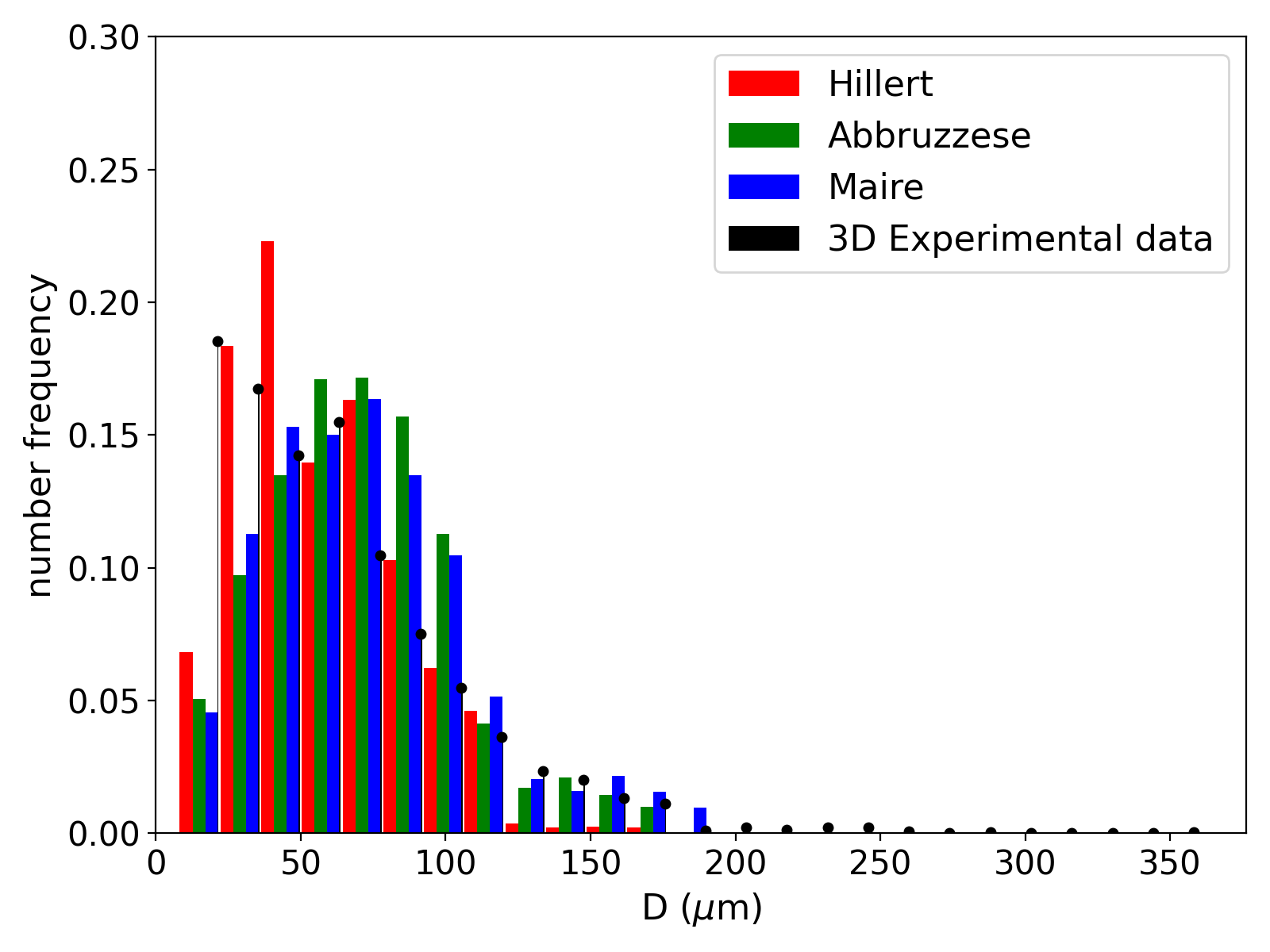}
    			\caption{\label{1100_2h_diam} 2h at at \SI{1100}{\celsius}.}
    		\end{subfigure}
          \begin{subfigure}[b]{0.35\textwidth}
        			\includegraphics[width=\textwidth]{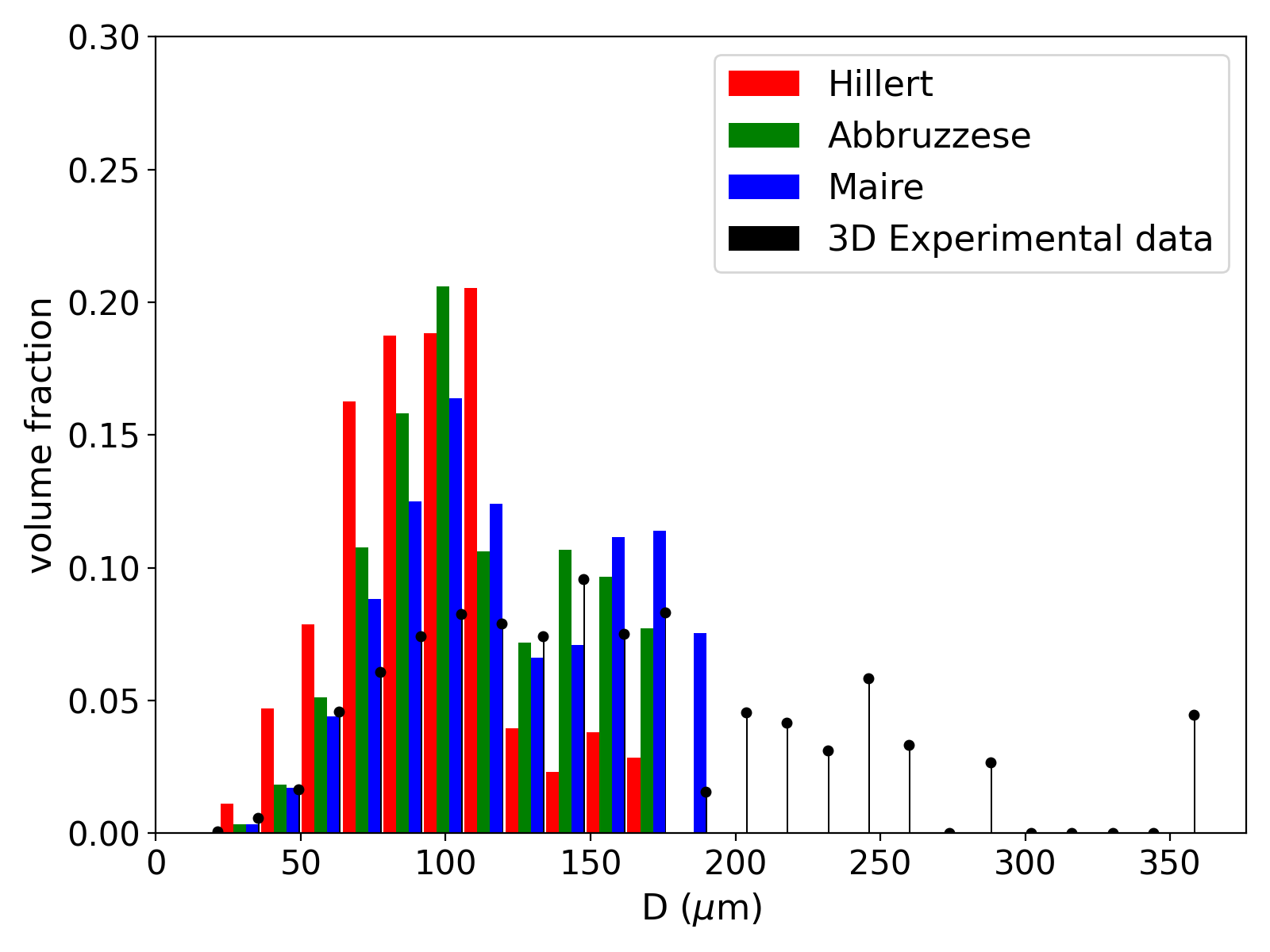}
        			\caption{\label{1100_2h_diam_surf} 2h at \SI{1100}{\celsius}.}
        		\end{subfigure}
          
		\begin{subfigure}[b]{0.35\textwidth}
			\includegraphics[width=\textwidth]{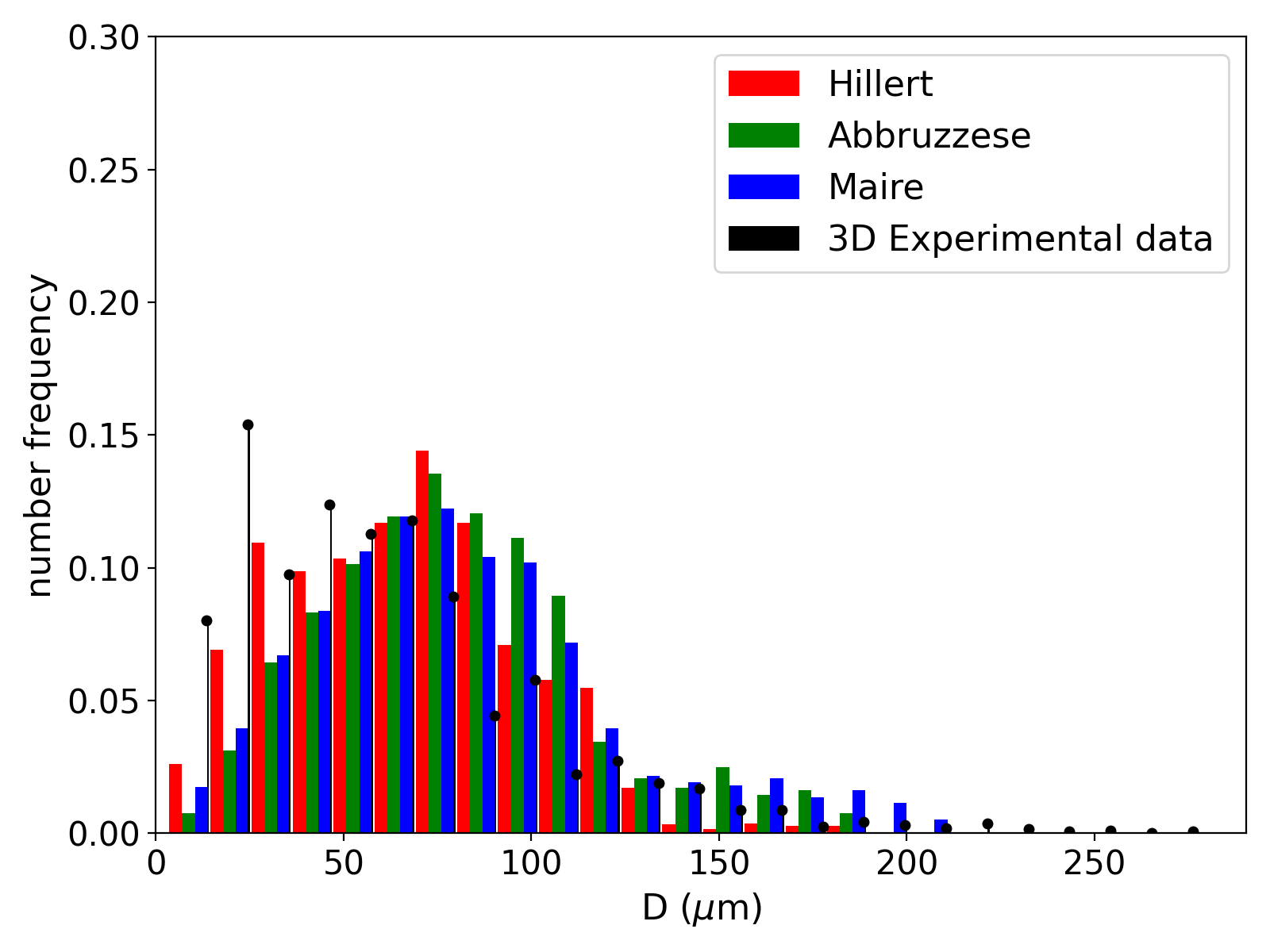}
			\caption{\label{1100_surf_3h_diam} 3h at \SI{1100}{\celsius}.}
		\end{subfigure}
          \begin{subfigure}[b]{0.35\textwidth}
        			\includegraphics[width=\textwidth]{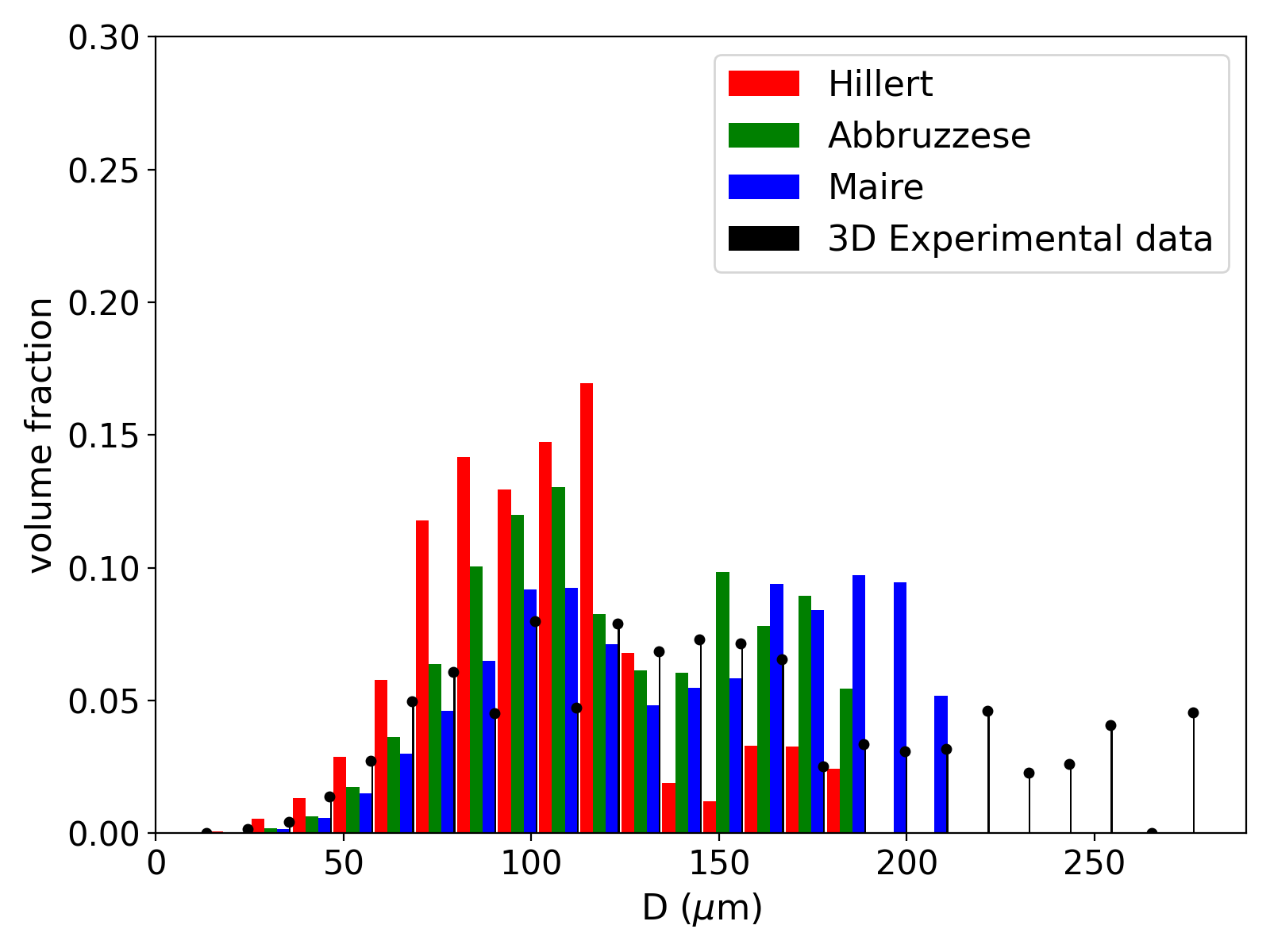}
        			\caption{\label{1100_surf_3h_diam_surf}3h at \SI{1100}{\celsius}}
        		\end{subfigure}
          
    		\begin{subfigure}[b]{0.35\textwidth}
    			\includegraphics[width=\textwidth]{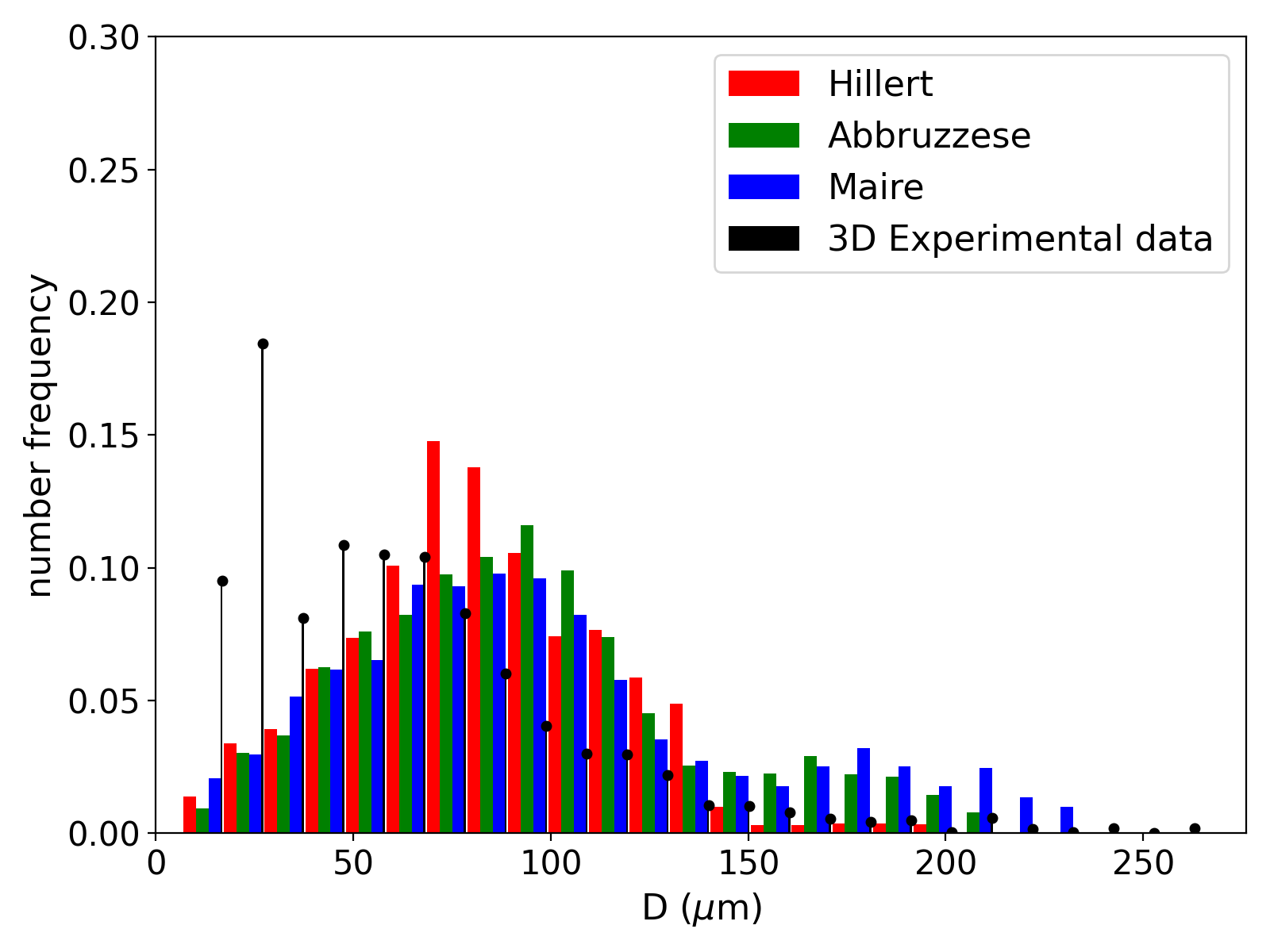}
    			\caption{\label{1100_surf_5h_diam} 5h at \SI{1100}{\celsius}.}
    		\end{subfigure}
		\begin{subfigure}[b]{0.35\textwidth}
			\includegraphics[width=\textwidth]{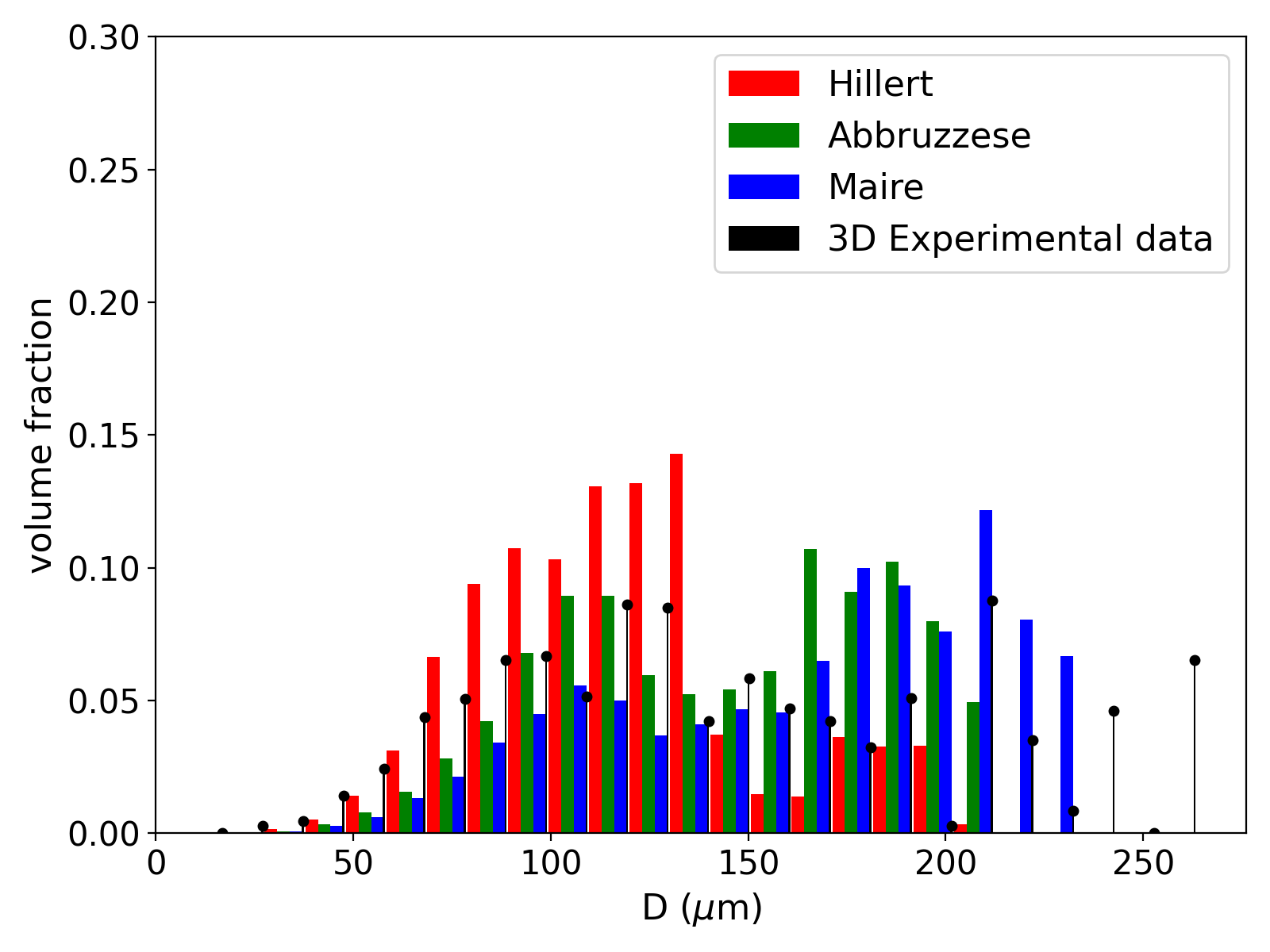}
			\caption{\label{1100_surf_5h_diam_surf} 5h at \SI{1100}{\celsius}}
		\end{subfigure}
  
		\caption{GSDs comparisons in (a,c,e,g) number frequency and (b,d,f,h) in volume fraction for Hillert, Abbruzzese  \textit{et al.} and Maire  \textit{et al.} models with experimental data at \SI{1100}{\celsius} for (a,b) 1h, (c,d) 2h, (e,f) 3h and (g,h) 5h of thermal treatment on 316L.}
        \label{fig:homo_asc}
	\end{figure}

Fig. \ref{fig:homo_asc} (a) to (h) illustrate mean-field GSDs predictions after different annealing times at \SI{1100}{\celsius} in comparison with experimental data obtained thanks to EBSD. The tail of GSDs representing larger grains is better predicted by Maire  \textit{et al.} model either considering frequency in number or volume fraction representation of the GSD. However, the volume fraction histograms show that none of the three models catches the entire experimental distribution tail for any of the studied annealing times. This may be due to both implementation hypotheses or also to experimental GSD statistical representativity. Hillert model seems to better predict number frequency histogram and under-estimate volumic predictions. However, for Abbruzzese \textit{et al.} and Maire \textit{et al.} models, the first part of the GSD, composed of small grain classes, shows good accordance with experimental data in volume fraction (fig. \ref{fig:homo_asc} (a), (c), (e), (g)). On the other hand, the number frequency predictions exposes some divergence with respect to the experimental data for small grain sizes. 

 One common strong hypothesis in models is considering spherical grains and spherical evolution with capillarity. Models also make an important approximation by considering grain boundaries properties as isotropic. Indeed, grain boundaries energy $\gamma_{GB}$ and their mobility $M_{GB}$ are both considered constant and identical for all grain classes used in the simulation. However, experimental microstructures have been proven \cite{rohrer_introduction_2010,kohara_anisotropy_1958}, to expose a dispersion in values of these properties. These hypotheses tend to smooth all microstructure heterogeneities at the beginning or during simulations, which may explains GSD extremity differences. The radius variation exchange considered in the GB migration equation for the different microstructure descriptions also impact the GSD prediction. Indeed, Hillert model by employing an HEM account for these variations through the choice of the MGS equation. 
 
 Experimental factors can also play a role in these differences. The statistical representativity of experimental data can be limited and the inverse Saltykov method is not a deterministic one. As a result no perfect reproducibility of the data is possible. This can explain the difficulty of models to predict GSD tail. The discussed models provide a general tendency of the GSD evolution depending on the thermal conditions. Moreover, no special treatment has been done to consider twining in this work even if these special GB can be considered as partially taken into account in identified parameters of the models such as $M_{GB}$.

\begin{figure}[!h]
		\centering
		\begin{subfigure}[b]{0.49\textwidth}
			\includegraphics[width=\textwidth]{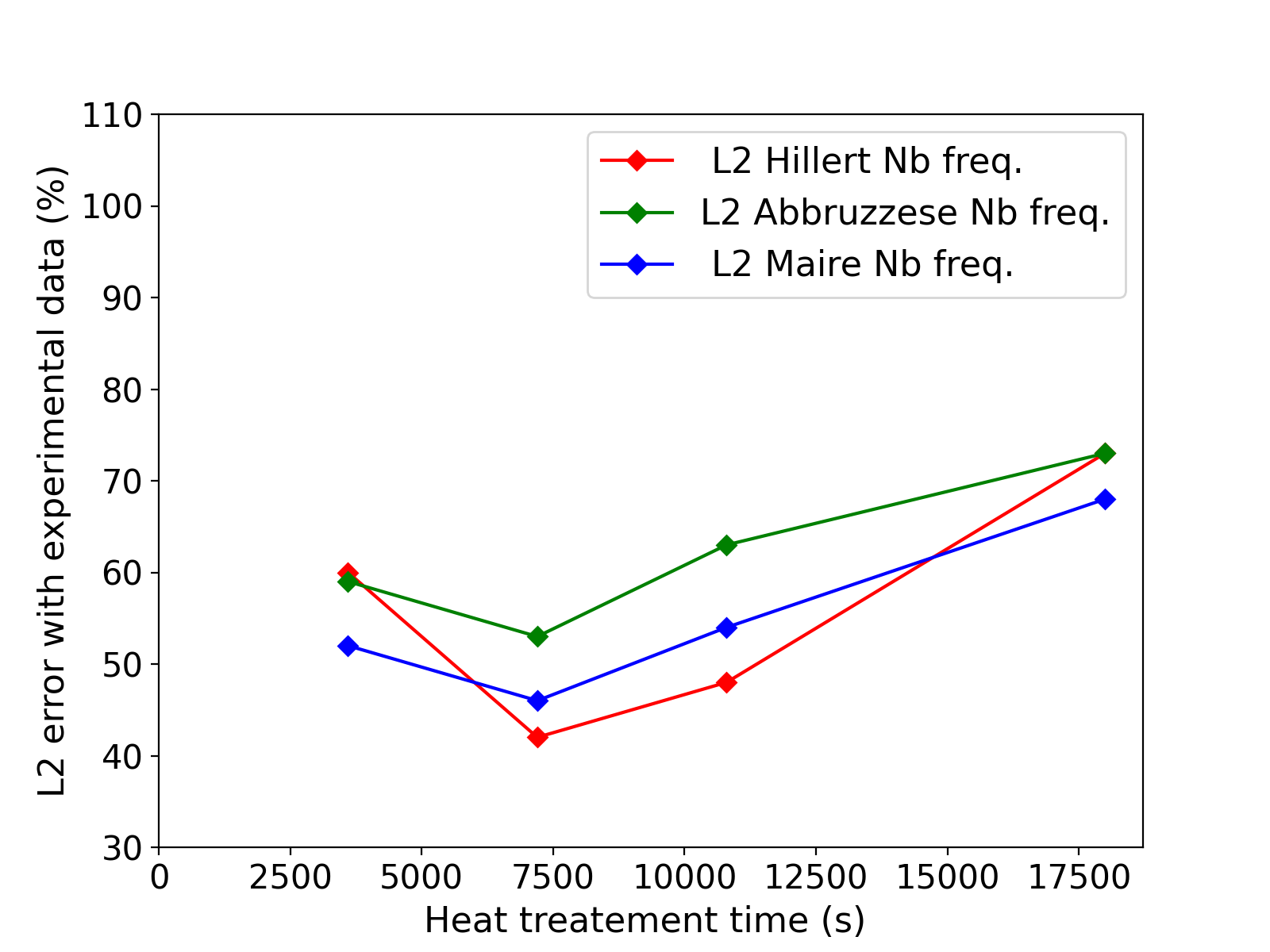}
			\caption{\label{L2_nb}  }
		\end{subfigure}
		\begin{subfigure}[b]{0.49\textwidth}
			\includegraphics[width=\textwidth]{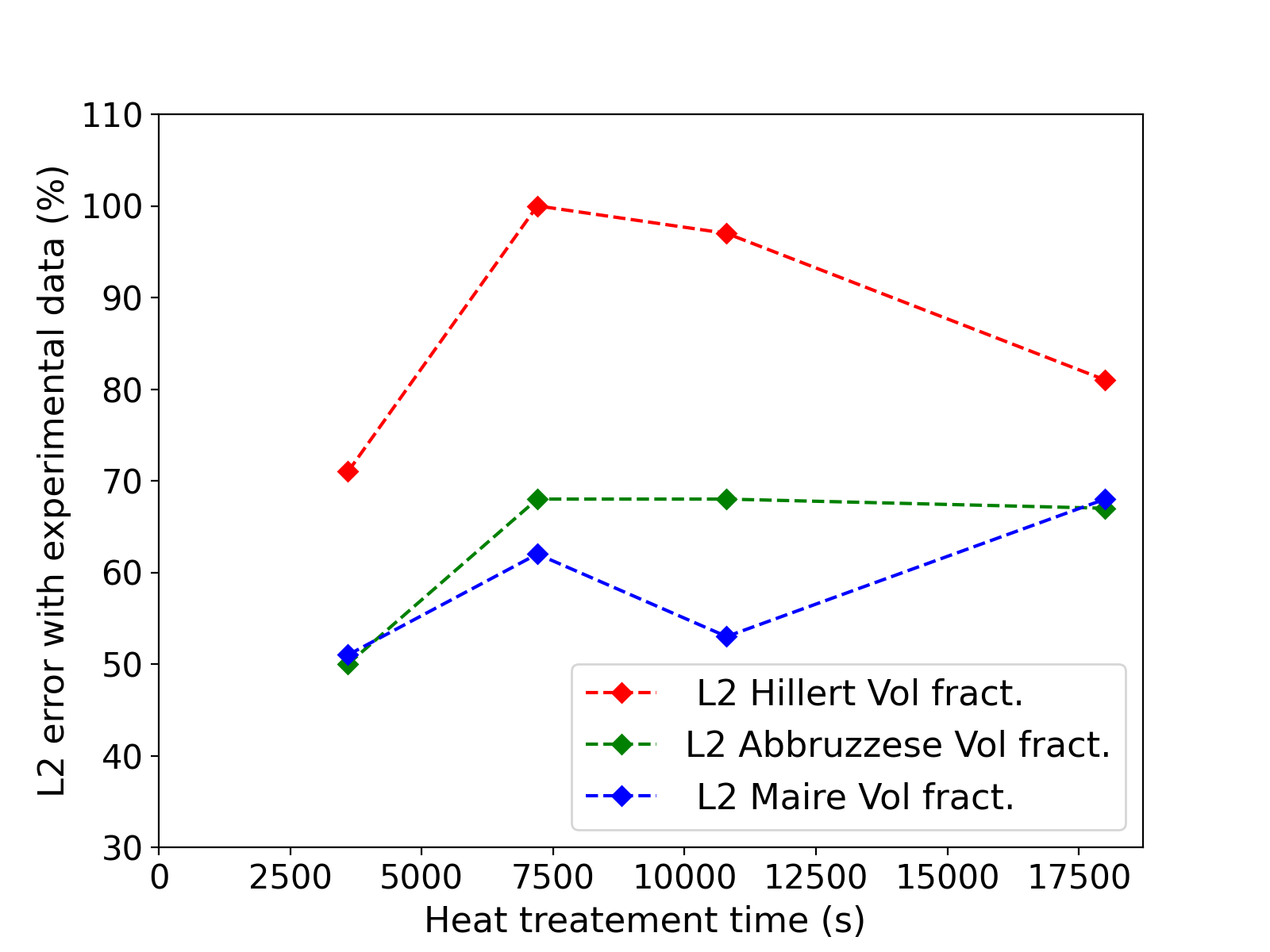}
			\caption{\label{L2_vol}}
		\end{subfigure}
        \caption{ \label{fig:hetero}Comparison of $L^2(t)$ error computed from GSDs for the three models and heat treatments from \SI{1}{\hour} to \SI{5}{\hour} at \SI{1100}{\celsius} with a description in (a) number frequency and (b) volume fraction.}
	\end{figure}

For more quantified comparisons, $L^2(t)$ error is computed with eq. \eqref{eq:L2}, with $S_i$ (resp.  $S_i'$) corresponding to the $i^{th}$ bin grain class area of the simulated GSD at time $t$ (resp. of the experimental GSD at time $t$). Fig. \ref{L2_nb} and \ref{L2_vol} illustrate the comparison of the latter criterion for the four experimented times in both representations in number and volume fraction. High $L^2(t)$ error values can be explained by the assumptions and experimental statistical representativity detailed above. However, these results provide a good relative comparison basis between models. As expected, Hillert model provide a good prediction rate in number frequency. But, its volumic predictions are low compared to the two others. Abbruzzese \textit{et al.} and Maire \textit{et al.} models provide a relatively constant prediction in both representations. In order to qualify a model prediction ability, good results in both representation are necessary. Maire \textit{et al.} expose here the overall lowest $L^2(t)$ values, making it the model with the highest accuracy compare to the others. 

In the case of GG, the simulation of an initial monomodal GSD is overall similarly predicted by the three models. Maire \textit{et al.} model improves GSD predictions on larger grain sizes but Hillert gives satisfying results considering the simplicity of its description when only GG mechanism is at play.

\subsubsection{Comparison of mean-field models on bimodal initial microstructure} \label{bi-mod distrib}

\begin{figure}[!h]
		\centering
		\begin{subfigure}[b]{0.49\textwidth}
			\includegraphics[width=\textwidth]{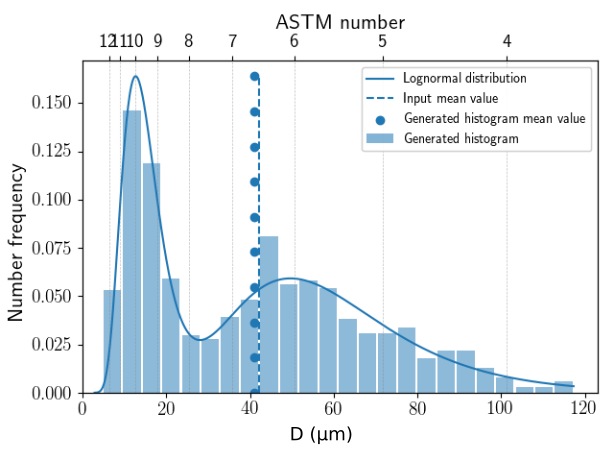}
			\caption{\label{bi_modal_distrib}}
		\end{subfigure}
		\begin{subfigure}[b]{0.49\textwidth}
			\includegraphics[width=\textwidth]{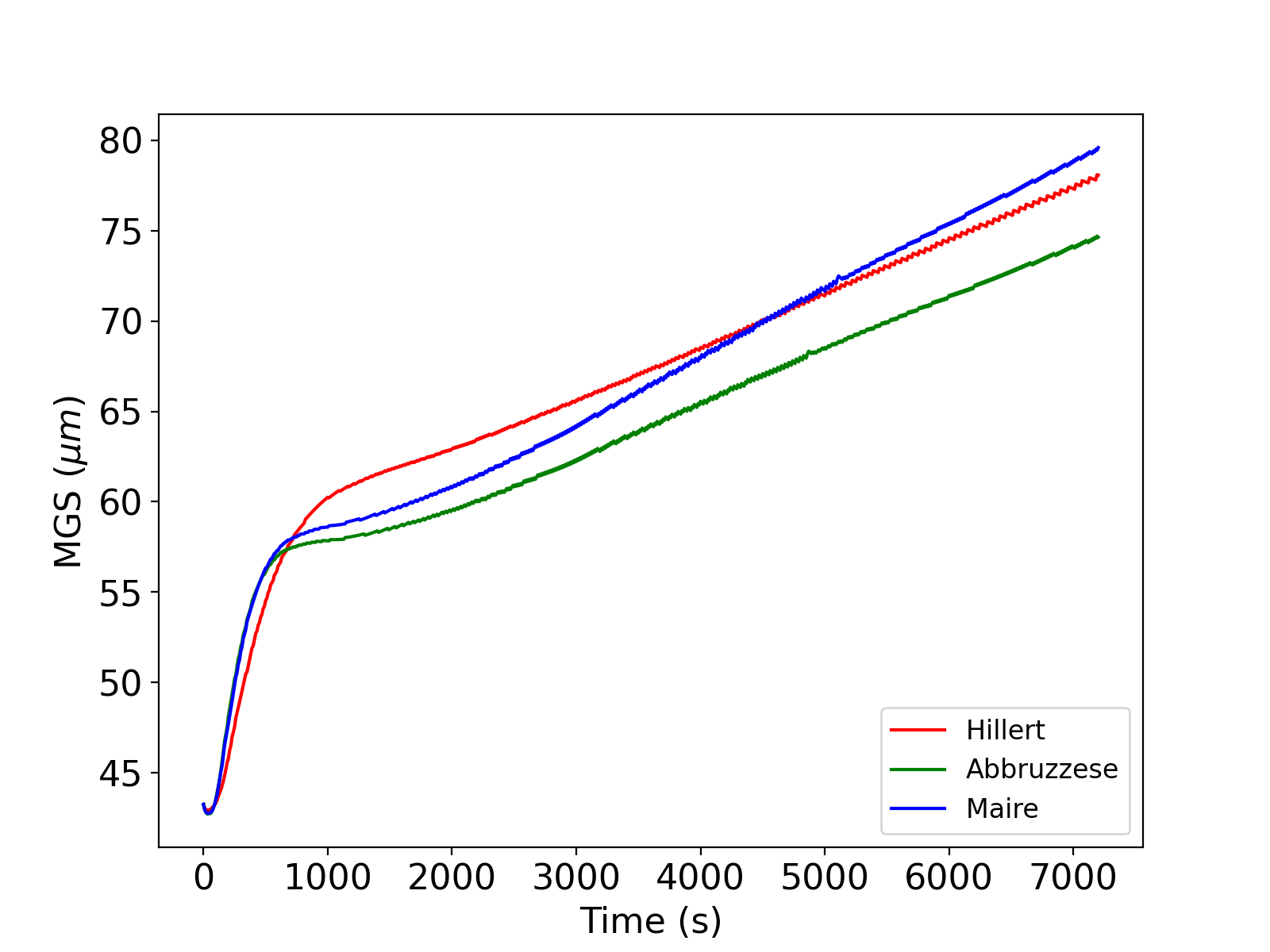}
			\caption{\label{mean_hetero} }
		\end{subfigure}
		\begin{subfigure}[b]{0.49\textwidth}
			\includegraphics[width=\textwidth]{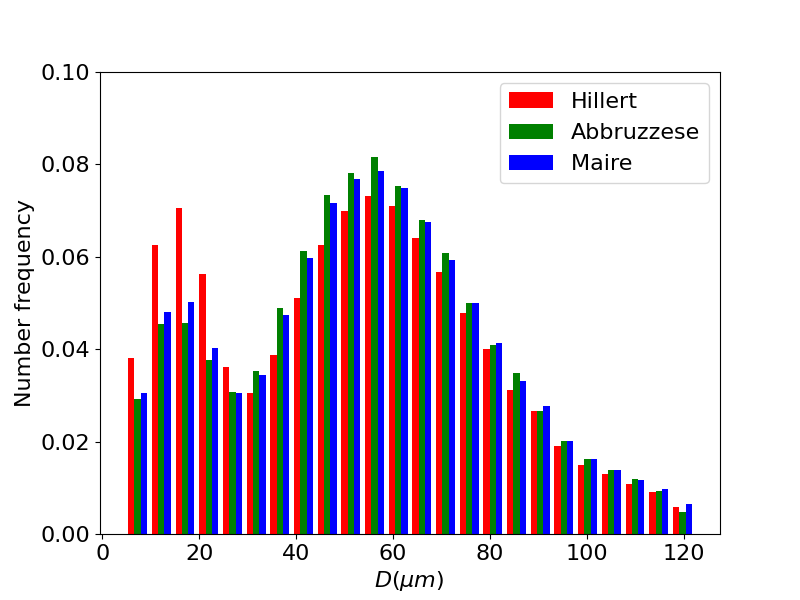}
			\caption{\label{GSD_hetero_5min}}
		\end{subfigure}
		\begin{subfigure}[b]{0.49\textwidth}
			\includegraphics[width=\textwidth]{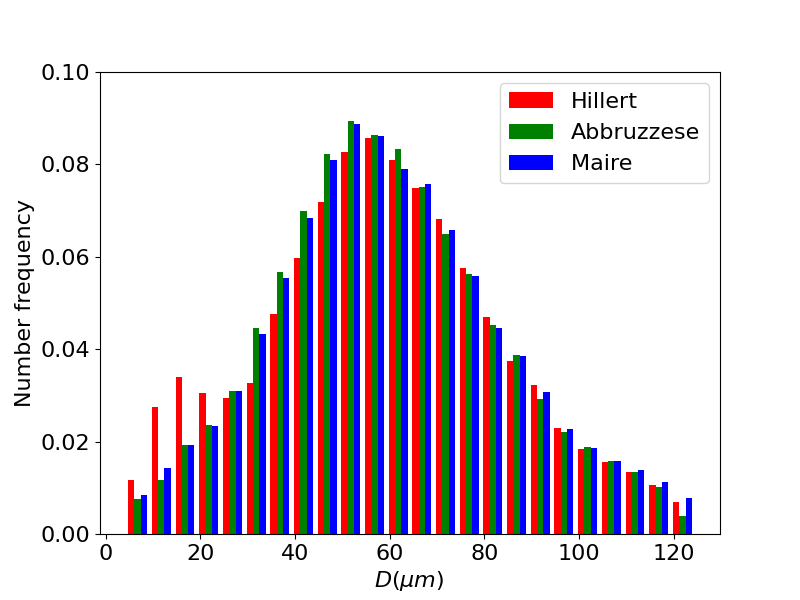}
			\caption{\label{GSD_hetero_10min}}
		\end{subfigure}
			\begin{subfigure}[b]{0.49\textwidth}
			\includegraphics[width=\textwidth]{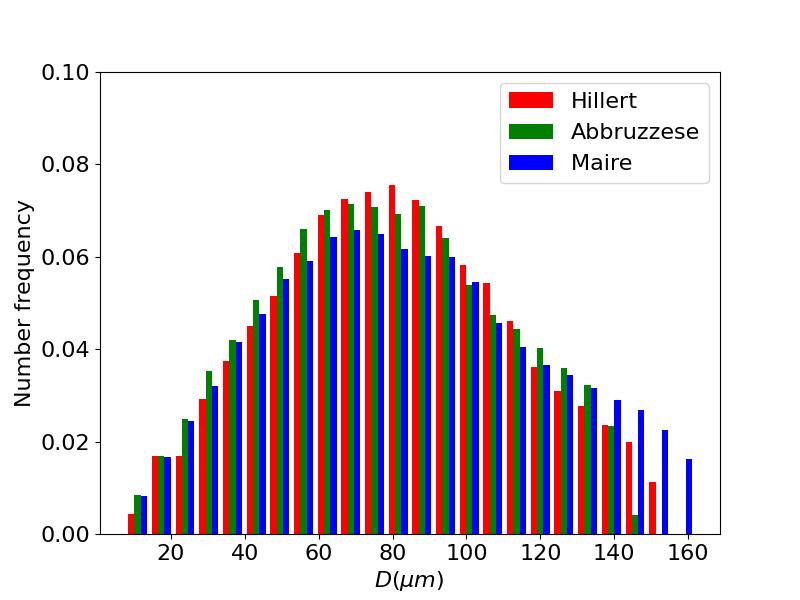}
			\caption{\label{GSD_hetero_2h}}
		\end{subfigure}
		\caption{Comparison of the GG predictions at \SI{1100}{\celsius} starting from an initial bimodal distribution with a neighborhood construction selected in the ascending order for the Maire \textit{et al.} model:  (a) Initial bimodal distribution, (b) MGS evolution in time, and GSDs at (c)  \SI{5}{\min}, (d) \SI{10}{\min} and (e) \SI{2}{\hour}.}
	\end{figure}

 By developing a test case using a bimodal initial GSD, the neighborhood construction of the Maire \textit{et al.} model is then tested for an initial heterogeneous GSD. An input microstructure is selected with a MGS of \SI{42}{\micro\meter} and a four ATSM grain size difference between the two selected grain populations to respect the bimodal definition of the ASTM standard E112 \cite{ASTM_E112} as shown in fig. \ref{bi_modal_distrib}. The previously identified reduced mobility values are used. In this case, no experimental data are available, thus only a relative comparison between models will be made. The MGS evolution can be visually dissociated into two kinetics. The first one is appearing in the range of the first \SI{10}{\min} then a more steady increasing kinetic until the end of the annealing treatment of \SI{2}{\hour} is observed in fig. \ref{mean_hetero}. Three associated GSD at \SI{5}{\min}, \SI{10}{\min}, and \SI{2}{\hour} comparing the behavior of the models are given in fig. \ref{GSD_hetero_5min}, \ref{GSD_hetero_10min} and \ref{GSD_hetero_2h}. For all cases, there are very few distinctions between models predictions and the bimodal aspect is rapidly smoothed after \SI{5}{\min} with the apperance of a monomodal distribution, excepted Hillert model predictions that retained a small bimodal distribution at \SI{5}{\min} and \SI{10}{\min}. The breaking point between the two kinetics in the MGS evolution corresponds to the point where the monomodal distribution is reached after about \SI{10}{\min}.  

	\label{fig:hetero}

	\paragraph{Impact of the selecting order of neighborhood construction on the distribution prediction}
	
\begin{figure}[!h]
		\centering
		\begin{subfigure}[b]{0.49\textwidth}
			\includegraphics[width=\textwidth]{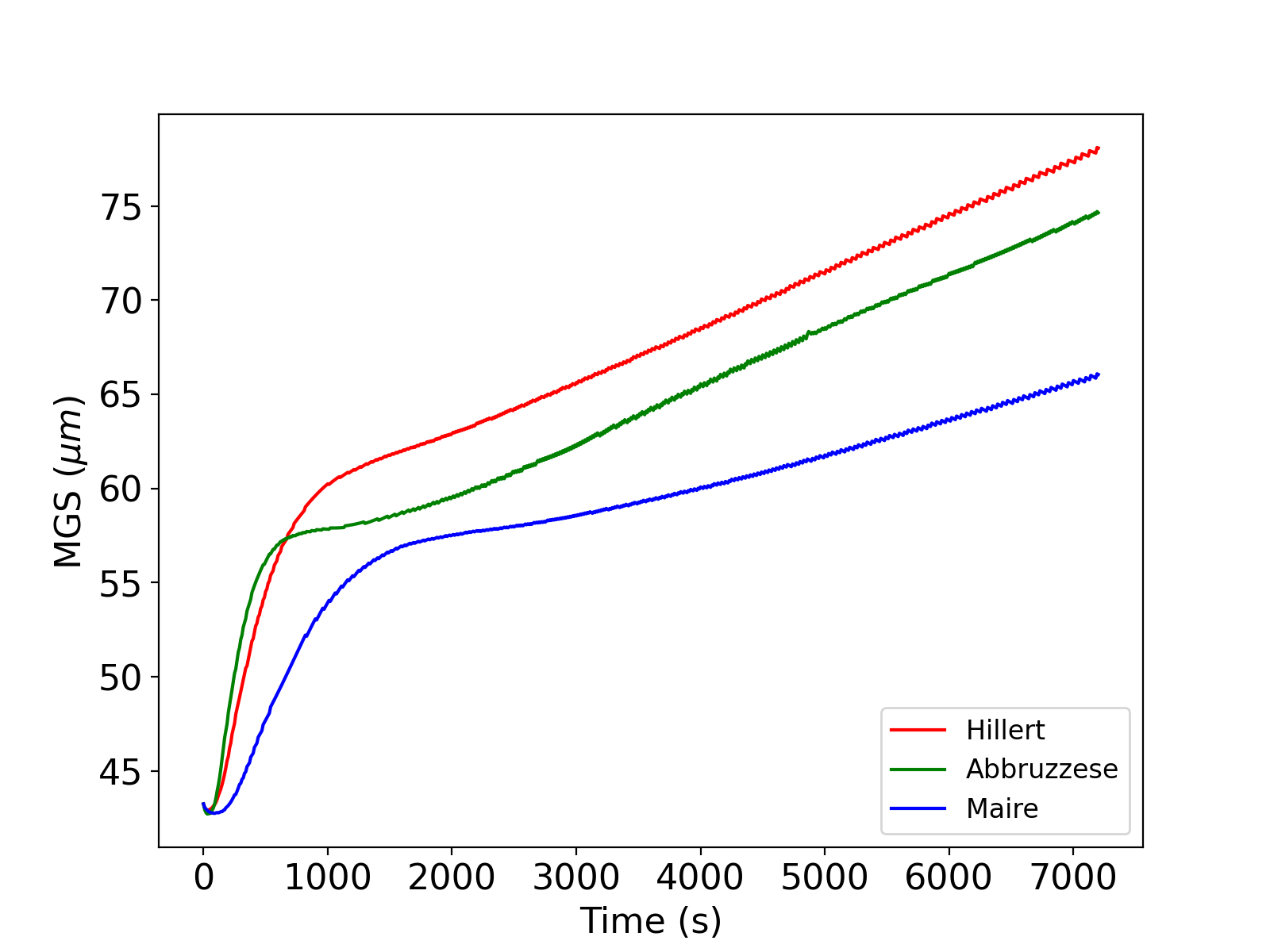}
			\caption{\label{mean_hetero_descending} }
		\end{subfigure}
		\begin{subfigure}[b]{0.49\textwidth}
			\includegraphics[width=\textwidth]{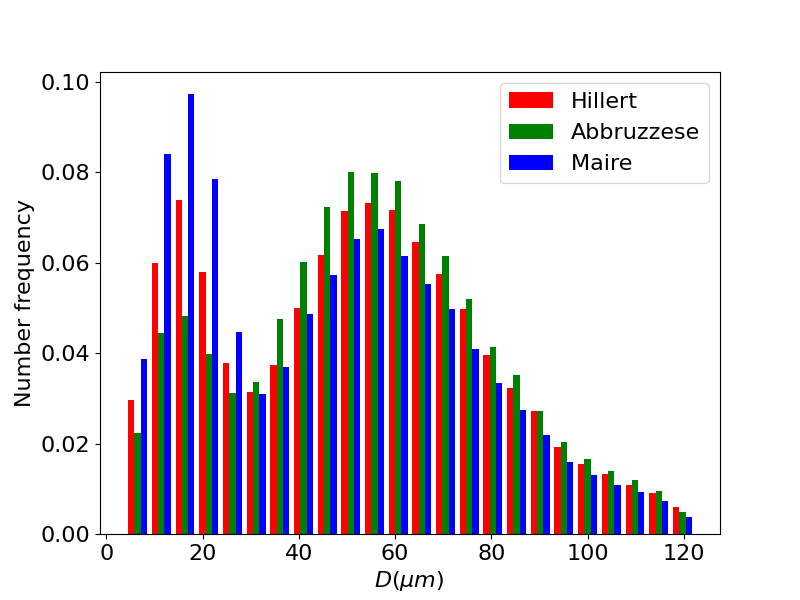}
			\caption{\label{GSD_hetero_desc_5min} }
		\end{subfigure}
		\begin{subfigure}[b]{0.49\textwidth}
			\includegraphics[width=\textwidth]{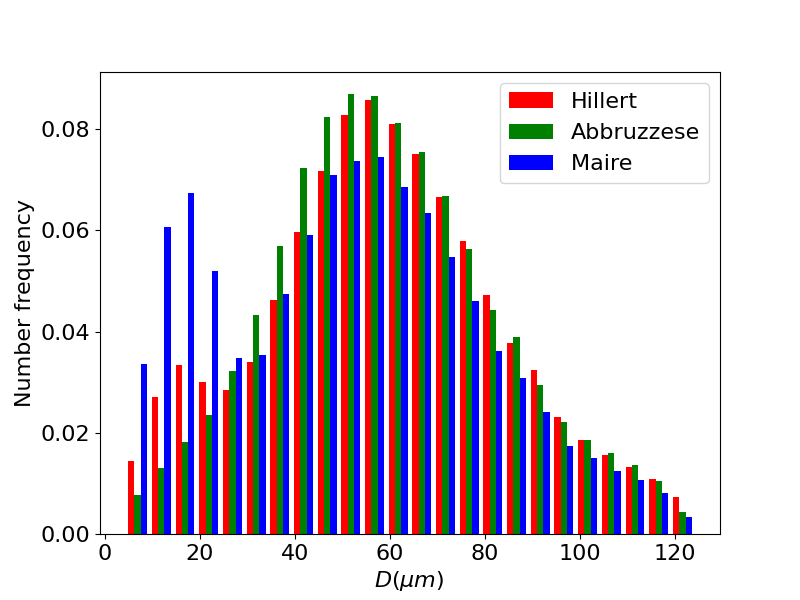}
			\caption{\label{GSD_hetero_desc_10min} }
		\end{subfigure}
		\begin{subfigure}[b]{0.49\textwidth}
			\includegraphics[width=\textwidth]{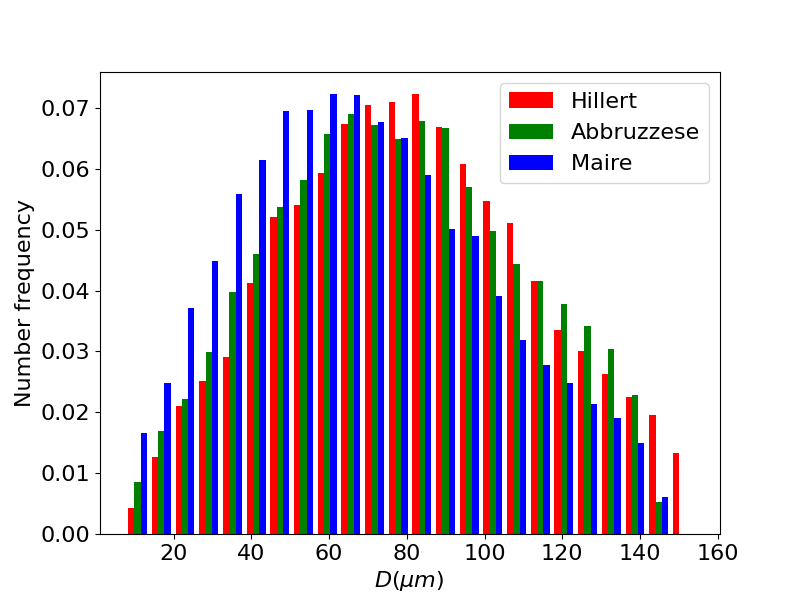}
			\caption{\label{GSD_hetero_desc_2h} }
		\end{subfigure}
		\caption{Comparison of the GG predictions at \SI{1100}{\celsius} starting from an initial bimodal distribution with a neighborhood construction selected in the descending order for the Maire \textit{et al.} model: (a) MGS evolution in time, and GSDs at (b) \SI{5}{\min}, (c) \SI{10}{\min} and (d) \SI{2}{\hour}.}
	\end{figure}

In the same way, as in section \ref{homo_descending}, an observation of the use of a descending selecting order of neighborhood construction is performed. Similarly, the MGS evolution of Hillert and Abbruzzese  \textit{et al.} in fig. \ref{mean_hetero_descending} remains unchanged and Maire\textit{ et al.} model kinetics is one more time slowed down. However, the associated GSDs at \SI{5}{\min} and \SI{10}{\min} show well the bimodality. Maire \textit{et al.} model seems to retard the smoothing effect conducting to the final Gaussian distribution as shown in fig. \ref{GSD_hetero_desc_5min} and \ref{GSD_hetero_desc_10min}. Indeed, in both graphs, the small grain size population keeps a higher number frequency in the case of Maire  \textit{et al.} model than for other GG models. In addition to the construction impact due to the selecting order as described in fig. \ref{contruction_oder}, this construction technique does not allow the same neighborhood diversity for all grain classes. As exposed by fig. \ref{Neighbor_construction}, grains classes selected in the middle of the construction process will benefit from a more important number of neighbors as part of it is built with blue grains having incomplete neighborhoods. As it gets to this end of the construction, less grain classes are available to be part of the neighborhood. Considering this point, the order of selection of neighborhood construction has an impact on the GSD prediction even if the bijectivity is taken into account. This topological information is consistent with what is observed on the GSD evolution.  In the bimodal case, it seems that the ascending order construction promotes the elimination of the small grains population of the distribution with a faster kinetic. Indeed, the curvature driving pressure of the GG phenomenon will tend to facilitate the disappearance of smaller grains considering the latter neighborhood construction. However, this elimination kinetic seems to be postponed when the descending construction order distribution is employed. In this case, small grains classes will be better dispersed in the neighborhood construction and least favorable to disappearance. This conducts to a higher MGS in the case of the ascending order distribution around \SI{80}{\micro\meter} while the descending counterpart case barely reaches \SI{66}{\micro\meter}. And finally, if an analysis of the remaining grain classes number at the end of the first phase around \SI{1000}{\second} of heat treatment is done, a difference of 90 grain classes in the considered system is observed. Sorting order of the grain classes using the specific neighborhood construction of the Maire \textit{et al.} model has a direct impact on microstructure evolution and therefore GSD predictions.

	\label{fig:hetero}
 
\section{Conclusion}
	
Different GG mean-field models were investigated in this article by first a detailed explanation of their equations/hypotheses and by comparing heat treatment predictions on 316L steel. The neighborhood description in Maire \textit{et al.} model is based on the original work of Abbruzzese \textit{et al.} \cite{Abbruzzese1992} that described a 2D GG model with a statistical neighborhood based on contact probabilities involving the entire microstructure. It relies on a hybrid description by using the statistical approach of contact probabilities to define neighbors surfaces in contact with grain classes coupled with a deterministic number of neighbors ruled by the use of a bijectivity criterion.  \\
To optimize the accuracy of the discussed models, parameters analyses were performed to observe their impact on the GSD predictions. First, a convergence study was achieved to optimize the initial and final number of grain classes in order to ensure statistical representativity. Then, the contact probability definition was identified to be a leverage quantity in the GSD description. Indeed, the latter impacts the distribution of the contact surface of neighbor classes. Four contact probability have been tested, from a ratio in number not involving the size of the grains to a volumic fraction. This means that for a contact probability in number, the neighbor grains size will only impact the grain evolution through its curvature radius in the GB migration. However, if a volumic contact probability is considered, a more important weight will be given to large grains which emphasizes the effect of grain topology. The volumic contact probability has been selected for the previous reasons. In particular, this has proven to give better GSD predictions by improving the description of the tail of the GSD when compared to experimental data as shown in section \ref{results}. And finally, the selected neighborhood construction order in the GSD, impacts strongly the microstructure evolution in Maire \textit{et al.} model. For this model, reduced mobility values need to be re-identified in order to be predictive when a different selecting order is adopted. 

Once optimized parameters were determined, a focus was made on comparing predicted GSD to EBSD experimental data for one to five hours annealing. To enable the comparisons, the identification of reduced mobility with respect to temperature and the use of the Saltykov method were necessary. Identified values of the reduced mobility ensure that the models exhibit a similar MGS evolution than the reference experimental data. On the other hand, Saltykov method enables to convert 2D GSD histograms to 3D discrete GSD providing a way for 2D to 3D conversion of experimental EBSD data. From these comparative histograms, a good general accordance of all models with experimental data was observed. However, Maire \textit{et al.} model gives more satisfying results in describing distribution tails than Hillert and Abbruzzese  \textit{et al.} ones. $L^2(t)$ computation also gives a reduced error for Maire \textit{et al.} model on the studied GG cases. When comparing implementation simplicity and GSD response, Hillert model gives good predictions with a simpler medium description.  \\

We have discussed in this article the interest of Maire \textit{et al.} model solely in the frame of GG. Originally designed to model discontinuous dynamic recrystallization phenomenon \cite{Maire2018}, its strength relies on accounting for topology in the considered grain neighborhoods. It is especially powerful when dealing with different types of grains. By distinguishing RX grains from non-RX ones, the specific neighborhood construction provides better predictions of GSD with respect to experimental GSD. While this article was dedicated to validate the neighborhood construction in the case of GG, new developments aiming to improve GSD predictions during and after recrystallization is to be followed.
	
\section*{Acknowledgements}
The authors thank ArcelorMittal, Aperam, Aubert \& Duval, CEA, Constellium, Framatome, Safran and Transvalor companies and the ANR for their support
through the DIGIMU consortium and RealIMotion ANR industrial Chair (Grant No. ANR-22-CHIN-0003).
	
	\bibliographystyle{unsrt}
	\bibliography{biblio}
	
\end{document}